%% file: main.tex
\documentclass[conference]{IEEEtran}
\usepackage{cite}
\usepackage{amsmath,amssymb,amsfonts}
\usepackage{algorithmic}
\usepackage{graphicx}
\usepackage{textcomp}
\usepackage{xcolor}
\usepackage{fancyhdr}
\usepackage[hyphens]{url}

\usepackage{soul}
\usepackage{algorithm}
\usepackage{xspace}
\usepackage[font=footnotesize,labelfont=bf]{caption}
\usepackage{mathtools}
\usepackage{xcolor}
\usepackage{amsfonts}
\usepackage{amssymb}
\usepackage{pifont}
\usepackage{color}

\usepackage{textgreek}
\usepackage{subfig}
\usepackage{textcomp}
\usepackage[]{siunitx}

\xspaceaddexceptions{]\}}
\newcommand{\ignore}[1]{}

\newcommand{\old}[1]{}
\newcommand{\fig}[1]{Figure~\ref{#1}}
\newcommand{\sect}[1]{Section~\ref{#1}}
\newcommand{\tab}[1]{Table~\ref{#1}}
\newcommand{\algo}[1]{Algorithm~\ref{#1}}
\newcommand{\eqn}[1]{Equation~\ref{#1}}

\newcommand{\lazyb}[0]{LazyBatching\xspace}

\newcommand{\infq}[0]{\texttt{InfQ}\xspace}
\newcommand{\batchT}[0]{\texttt{BatchTable}\xspace}

\newcommand\blfootnote[1]{%
\begingroup
\renewcommand\thefootnote{}\footnote{#1}%
\addtocounter{footnote}{-1}%
\endgroup
}

\renewcommand{\baselinestretch}{.999}
\title{\huge LazyBatching: An SLA-aware Batching System\\ for Cloud Machine Learning Inference}

\begin{document}

\author{
\IEEEauthorblockN{
Yujeong Choi\hspace{2.3em}Yunseong Kim\hspace{2.6em}Minsoo Rhu\hspace{1em}}
\IEEEauthorblockA{
School of Electrical Engineering\\
KAIST\\
\texttt{\{yjchoi0606, yskimno1, mrhu\}@kaist.ac.kr}\\
}
}


\maketitle
\pagestyle{plain}

\input{tex/abstract}

\IEEEpeerreviewmaketitle


\blfootnote{
This is the author preprint version of the work.
}

\input{tex/intro}

\input{tex/background}

\input{tex/motivation}

\input{tex/proposed}
\input{tex/methodology}

\input{tex/result}

\input{tex/discussion}

\input{tex/conclusion}


\bibliographystyle{IEEEtranS}
\bibliography{ref}

\end{document}

%% file: tex/abstract.tex
\begin{abstract}

In cloud ML inference systems, batching is an essential
	 technique to increase throughput which
	 helps optimize total-cost-of-ownership.  Prior graph batching combines
	 the individual DNN
	 graphs into a single one, allowing multiple inputs to be
	 concurrently executed in parallel.  We observe that the
	 coarse-grained graph batching becomes suboptimal in effectively handling the
	 dynamic inference request traffic, leaving significant performance left on
	 the table. This paper proposes \lazyb, an SLA-aware batching system that
	 considers both scheduling and batching in the granularity of individual
	 graph nodes, rather than the entire graph for flexible batching. We show that \lazyb can
	 intelligently determine the set of nodes that can be
	 efficiently batched together, achieving an average $15\times$,
		 $1.5\times$, and $5.5\times$ improvement than graph batching in terms of average response time,
		 throughput, and SLA satisfaction, respectively.

	\end{abstract}

%% file: tex/intro.tex
\section{Introduction}

As the demands for accelerating deep neural network (DNN) based machine
learning (ML) algorithms increase, several hyperscalers have begun offering the
compute and memory required for DNN training and inference as a service to
end-users using
off-the-shelf CPUs, GPUs, or custom-designed ML accelerators such as neural
processing units (NPUs)~\cite{google_cloud_ml,amazon_sagemaker,azure_ml}. While inference on the edge has recently received
significant attention in certain application domains, major IT vendors are
still predominantly deploying ML inference service over the cloud.  
As end-users typically desire real-time response, providing
low latency inference is a fundamental requirement in cloud ML systems.
However, achieving high resource utilization and system throughput is still
vital in these consolidated/virtualized warehouse-scale computers as it helps
optimize the total-cost-of-ownership.

Given this landscape, existing ML frameworks for \emph{serving} 
ML inference requests~\cite{tensorrt_inf_server,tf_serving}	provide support for \emph{batching} inputs. Batching is
an essential technique in ML frameworks to increase system throughput as it
better utilizes parallelism and locality across the batched inputs.  Current ML
frameworks typically express the DNN algorithm as a computation \emph{graph},
					 and batching is conducted in the entire graph granularity
					 (i.e., entire DNN). These so-called \emph{graph batching}
					 solutions combine the individual dataflow graphs into a single one,
					 which is concurrently executed by the backend processor in unison
					 for higher computational efficiency and
					 throughput~\cite{tensorrt_inf_server,cellular_batching}. As all training inputs
					 are available before the training begins, graph batching can
					 effectively collect multiple training inputs to form a batch
					 without any delays. However, batching inputs for inference is
					 non-trivial as the ML inference server receives inputs at different
					 times, the arrival rate of which is determined by the popularity of
					 the deployed model.  As such, graph batching
					 for inference must carefully balance the tradeoff between latency
					 and throughput. For instance, a large batch size might help improve
					 throughput but the scheduler must then wait for a longer period of
					 time to batch large enough inputs, suffering from an added latency.
					 An insufficient level of batching on the other hand can help reduce
					 latency, but comes at the cost of aggravated throughput.
					 Consequently, existing graph batching solutions provide
					 \emph{model-allowed maximum batch size} (i.e., the inference server
							 will schedule the batched input once a certain number of inputs
							 are collected) and \emph{batching time-window} (i.e., the
								 longest time period the inference server will wait for inputs
								 to form a batch) as hyperparameters of the inference server.  
							 Unfortunately, a {\bf key challenge} of graph batching
							 is that a statically configured batching system (with 
									 maximum batch size and batching time-window) must handle all
							 deployment scenarios which can be suboptimal depending on the inference request
							 traffic and available resources (\sect{sect:motivation}).  For instance, having a large batching
							 time-window is a	significant overkill under lightly loaded
							 inference request traffic because the requests queued inside the
							 server must needlessly wait during the batching time-window,
							 slowing down average response time.  Conversely, a large
							 batching time-window and batch size can be advantageous for
							 periods when the server is heavily congested. The
							 baseline graph batching however cannot flexibly adjust to the
							 dynamic server traffic, leaving significant performance left
							 on the table.

	 In this paper, we propose \emph{LazyBatching}, an intelligent batching system that
	 dynamically adjusts the level of batching to balance latency, throughput, and SLA (service level
			 agreement) satisfaction.  
	 A key limitation of conventional graph batching is its
	 inability to service newly arrived requests when an ongoing batch of
	 requests has yet to complete its execution.  Rather than having a batched
	 input execute uninterrupted until the entire graph completes, \lazyb maintains
	 the scheduling granularity in fine-grained \emph{node-level} (i.e., layer granularity) and allows
	 different (batched) inputs to be \emph{interleaved} for execution.  In
	 effect, our \lazyb scheduler can \emph{preempt} and stall the currently
	 ongoing batch until the newly arrived input is preferentially scheduled to
	 \emph{catch up} the progress of the preempted batches.  Such flexible node-level
	 scheduling enables	the	 preempted and preempting requests to be batched at any given layer,
	 which 
	 significantly improves the batching opportunities.
	 Naturally, the effectiveness of ``\emph{lazily}'' batching inputs as
	 outlined above is dependent on whether the preempted batch inputs are
	 still able to meet SLA goals despite having to wait for the newly received
	 inputs to catch up its progress. The {\bf key innovation} of our proposed \lazyb
	 is the development of an SLA-aware scheduling algorithm that utilizes
	 domain-specific properties of ML inference to intelligently decide
	 when/which inputs to lazily batch or not. Concretely, \lazyb determines
	 what is the remaining SLA \emph{slack time} of a currently ongoing request and
	 utilizes that information to dynamically judge whether to preempt or
	 continue that ongoing request to satisfy SLA goals while maximizing system
	 throughput. Because \lazyb can flexibly adapt its batching level for both heavily or lightly loaded
	 inference request traffic conditions, it liberates the end-user from searching
	 the optimal batching hyperparameters (e.g., batch time-window
			 and maximum batch size) as done in conventional, static graph batching.
	  In effect, our proposed \lazyb system helps improve throughput
	while still meeting the SLA goals of
	 cloud ML inference. Below we summarize our {\bf key contributions}:

\begin{itemize}

\item We develop an SLA-aware slack prediction model which exploits a domain-specific
property of ML inference to predict the DNN inference time
for slack estimation.

\item We propose \lazyb, a low-cost and practical batching system for cloud
inference. Unlike prior work, our solution is not limited to a particular type of a DNN layer
and can flexibly adapt to the deployment environment without hand-tuning the
batching parameters.
	
\item Compared to graph batching, \lazyb provides an average $15\times$, $1.5\times$, and $5.5\times$
improvement in latency, throughput, and SLA satisfaction, respectively.

\end{itemize}

%% file: tex/background.tex
\section{Background}
\label{sect:background}

\subsection{Deep Neural Networks}
\label{sect:dnn}

\begin{figure}[t!] \centering
\includegraphics[width=0.43\textwidth]{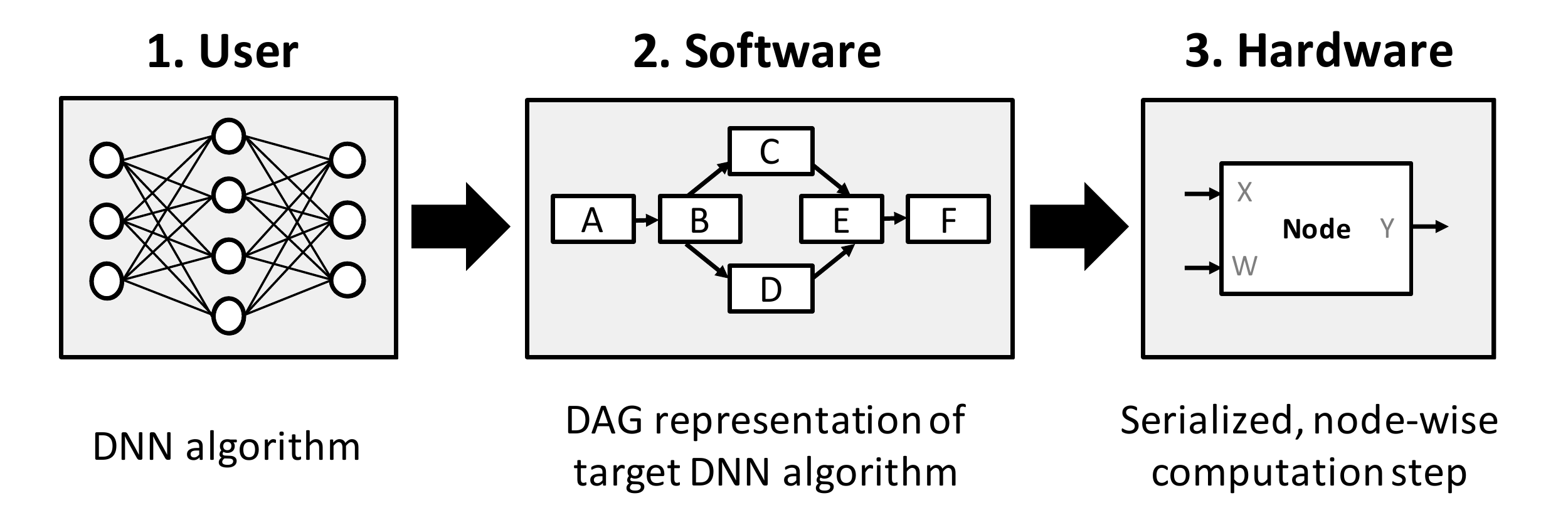}
\vspace{-0.5em}
\caption{
DNN execution flow, from a high-level ML framework down to low-level
hardware architecture. 
}
\vspace{-1.3em}
\label{fig:dnn_and_dag}
\end{figure}

\begin{figure}[t!] \centering
\includegraphics[width=0.485\textwidth]{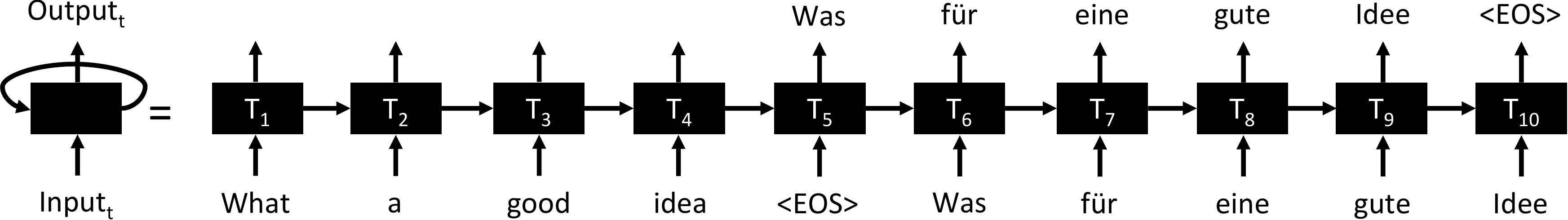}
\caption{
A \emph{vanilla} recurrent layer unrolled into ten sequence length 
in this English-to-German translation example.
For different input sentences,
		the output sequence length can be different (e.g., ``The sky is blue''$\rightarrow$``Der Himmel ist blau'').
The recursive, time-unrolling effects in attention-based	NLPs~\cite{transformer,bert,gpt2}
are observed in the decoder blocks of these algorithms.
}
\vspace{-1.3em}
\label{fig:dag}
\end{figure}

Deep neural network (DNN) based ML applications are represented as a
direct acyclic graph (DAG) in popular ML frameworks~\cite{tf,pytorch,mxnet}.
Each node within the DAG corresponds to a DNN layer, which is commonly designed
using convolutional, activation, pooling, fully-connected, and recurrent
layers.  \fig{fig:dnn_and_dag} shows how the DAG based DNN is lowered
into a serialized, node-wise (i.e., layer-wise) execution step.   ML applications for computer vision
are primarily based on convolutional neural networks (CNNs) using convolutional
and fully-connected layers. These DNN applications typically have a
\emph{static} graph topology where the number of nodes within the graph and its
structure are fixed (e.g., the DAG in \fig{fig:dnn_and_dag}).
In contrast, DNNs used for speech recognition~\cite{deepspeech2} or natural
language processing (NLP)~\cite{bert} exhibit a \emph{dynamic} graph structure
in that they have \emph{variable} number of graph nodes to traverse within the
DAG.  These applications are designed to model the so-called
``sequence-to-sequence'' (seq2seq) behavior: e.g., translating a English
sentence to German involves mapping a variable length sequence of English words
into a variable length sequence of German words~\cite{seq2seq}. As such, the
DNNs used to model the seq2seq behavior have \emph{recursive} computations,
		 rendering the graph topology of these DNNs to be dynamically derived in an
		 input-dependent manner.  In other words, the recursive computations within
		 these dynamic graphs must be \emph{unrolled} in order to accurately
		 reflect the seq2seq behavior (\fig{fig:dag}). Seq2seq models traditionally utilized recurrent neural networks (RNNs)
	using LSTM or GRU cells as building blocks for the recurrent
	layers~\cite{lstm,gru}.  Recent work however demonstrated that
	\emph{attention} modules (pioneered by the work on
			\emph{Transformers}~\cite{transformer}) can achieve superior algorithmic
	performance than RNNs (e.g., BERT~\cite{bert}, GPT-2~\cite{gpt2}).
	Consequently, state-of-the-art NLP applications are primarily designed using
	attentions these days.

\subsection{Batching for Training vs. Inference}

A DNN application must be \emph{trained} in order to be deployed for
\emph{inference}.  Batching is a critical component for both training and inference in today's ML
frameworks as it helps increase throughput and optimize the
total-cost-of-ownership in cloud ML systems. Because the training
dataset is already available before the learning process begins, constructing a
large enough batch size is trivial for training.
However, collecting batched inputs for inference is
challenging because the server receives DNN inference requests 
at varying rates, which is determined as a function of how popular the
deployed model is, what time of the day the requests are being received, and
etc.
In the rest of this
paper, we focus on inference which has unique challenges in
developing an effective batching system.

\subsection{Batching on Latency vs. Throughput}
\label{sect:background_batching}

\begin{figure}[t!] \centering
\includegraphics[width=0.485\textwidth]{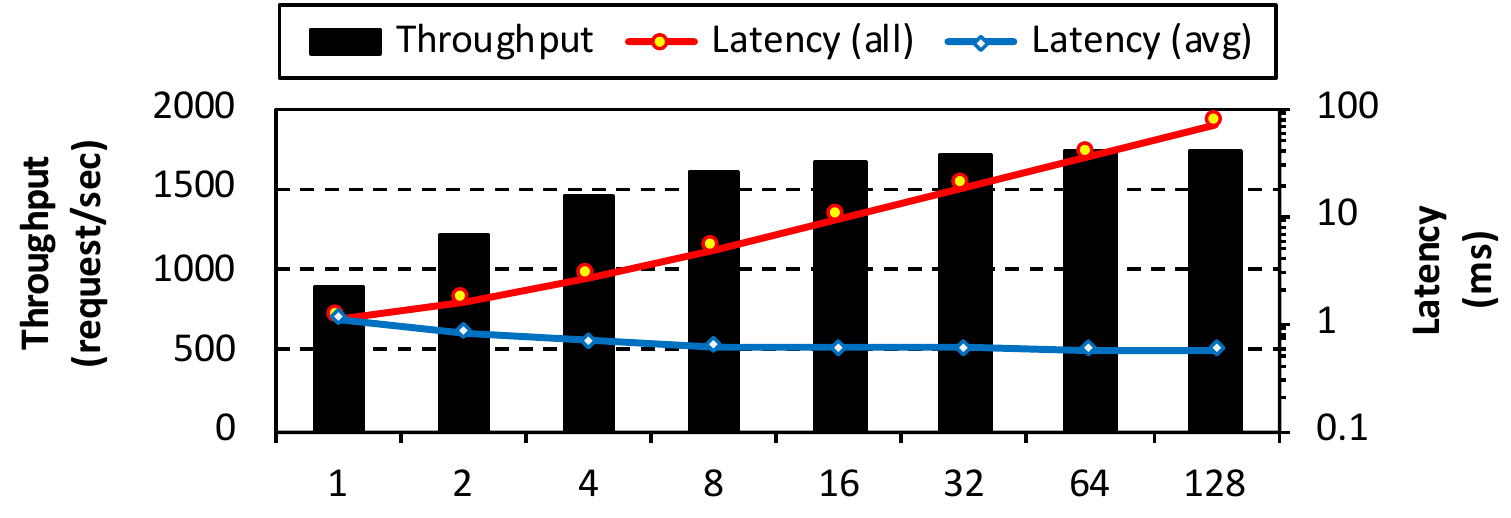}
\vspace{-0.5em}
\caption{
   Effect of batching on throughput (left-axis) and overall latency of batched execution (red, right-axis)
as a function	of batch size (x-axis). 
		To show the benefits of batched execution on reducing the average latency per each individual input,
			 the blue line represents average latency per each input (i.e., Latency(all)/number of batches) on
				 the right-axis.
		For this experiment, we assume that the batched inputs are already formed 
	at size $N$, without waiting for them to be collected. 
}
\vspace{-1.3em}
\label{fig:batching_impact}
\end{figure}

		Popular ML model serving frameworks such as TensorRT Inference
		Server~\cite{tensorrt_inf_server} or TensorFlow Serving~\cite{tf_serving} conduct batching in the
		entire \emph{graph} granularity.  These so-called \emph{graph batching} solutions
		combine multiple individual DAG into a single, batched DAG, allowing the
		backend processor to execute them in parallel.
		\fig{fig:batching_impact} shows the effect of graph-level batching in ResNet's effective throughput and latency.
		As depicted, the effective throughput rapidly increases as batch size
		gets larger, which amortizes the cost of inference and translates into a 
		sharp reduction in average inference
		latency per each input (the blue line, Latency(avg)). 
		This is because having a larger batch size 
		increases the required computations which helps better saturate the
				abundant computing resources within GPUs/NPUs for higher throughput. 
	However, the increase (decrease) in throughput (latency(avg))
	eventually levels out beyond a certain batch level,
	highlighting the importance of selecting an optimal level of batching that 
	balances throughput and latency.

	\subsection{Research Scope}
\label{sect:scope}

While throughput-optimized GPUs fit well for training, they are often deemed
ill-suited for latency-critical, low-batched inference because of their low
utilization~\cite{tpu1,facebook_hpca2018}.  Consequently, recent cloud ML
systems~\cite{tpu1,brainwave_isca,xilinx_versal,article_about_FB_developing_NPUs,article_about_Amazon_developing_NPUs}
employ custom designed NPUs for deployment (e.g., Google's TPU~\cite{tpu2},
		Habana's Goya~\cite{goya}, Facebook's Kings Canyon~\cite{kings_canyon}).
LazyBatching is applicable for both GPUs and NPUs, but given the
popularity of NPUs for latency-critical inference scenarios, 
we assume NPUs as the baseline accelerator
architecture in this paper.  Nonetheless, 
LazyBatching's effectiveness over GPU-based inference
systems is quantitatively demonstrated in \sect{sect:eval_sensitivity}.

%% file: tex/motivation.tex
\section{Motivation}
\label{sect:motivation}

\subsection{Limits of ``One-Size-Fits-All'' Batching}
\label{sect:limits_graph_batching}

\begin{figure}[t!] \centering
\includegraphics[width=0.45\textwidth]{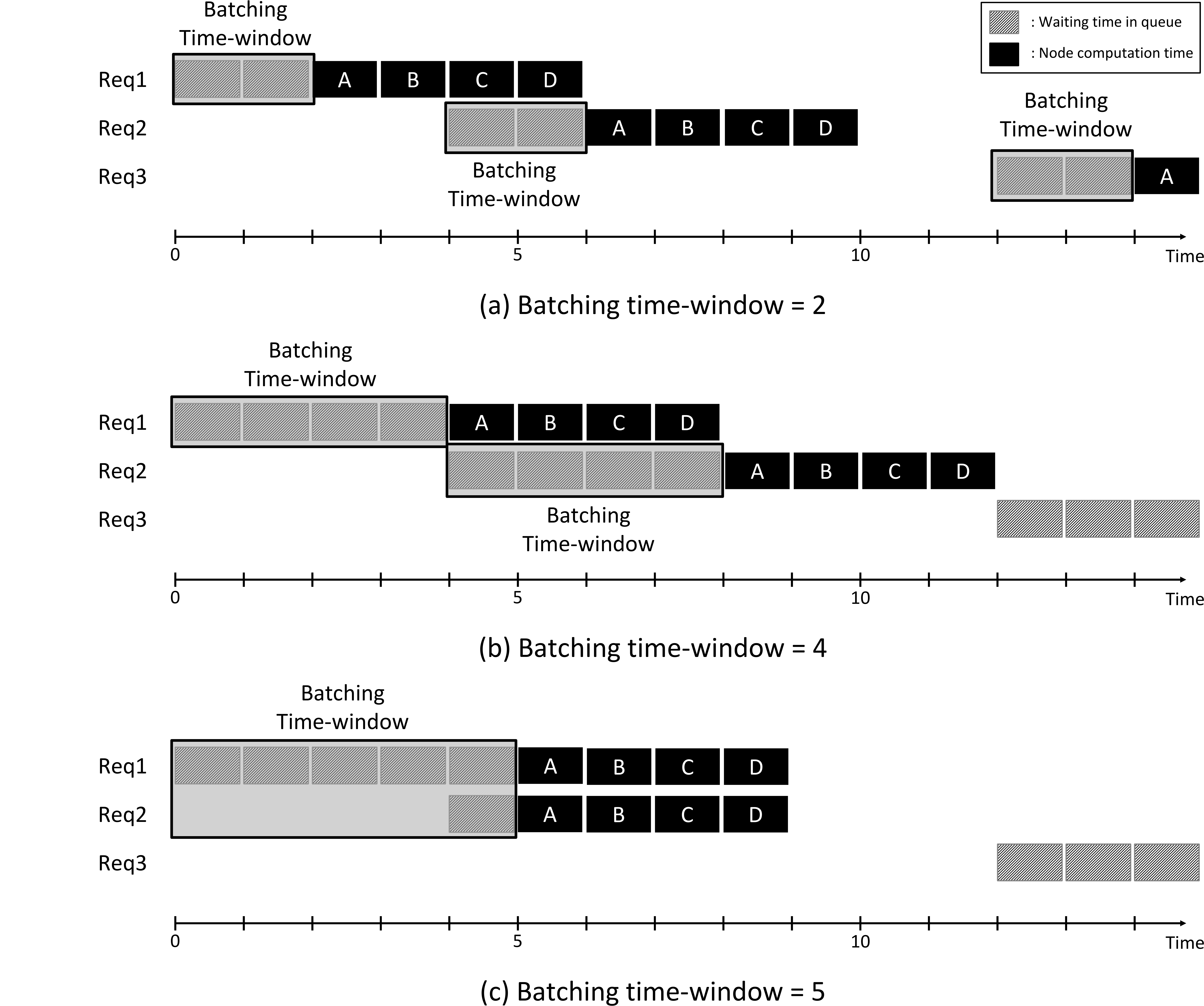}
\vspace{-0.3em}
\caption{
Timeline of baseline graph batching when the batching time-window is changed.
Example assumes the server receives $Req2$ and $Req3$ at t=$4$ and t=$12$,
respectively.
}
\vspace{-1.55em}
\label{fig:graph_batching_parameters}
\end{figure}

Conventional graph batching takes a ``one-size-fits-all'' approach,
						 which utilizes the following two hyperparameters to optimize the ML
						 inference server.  First, the \emph{model-allowed maximum batch
							 size} is used to configure the scheduler to only batch inputs up
							 to the point where having a larger batch size helps improve
							 throughput while still improving user-responsiveness.  In
							 \fig{fig:batching_impact} for instance, it is practically
							 meaningless for the  ML inference server to batch inputs beyond
							 $16$ for ResNet as the effective
							 throughput is saturated beyond this point.  Second, the ML
							 inference server is also setup with a \emph{batching
								 time-window} which is the maximum period of time the scheduler
								 waits for incoming requests to form a larger batch.  When
								 the request traffic to the inference server is lightly loaded,
								 having a smaller batching time-window prevents 
								 the server from needlessly waiting for future inputs to batch. 
								 For instance, increasing batching time-window from $2$
								 to $4$ in \fig{fig:graph_batching_parameters}(a-b) does not help
								 increase batch size and needlessly delay the time $Req1$ and
								 $Req2$ can start execution.  Conversely, when the inference
								 request traffic is high, this hyperparameter can help guarantee that
								 the server waits long enough to form a larger batch
								 to increase throughput while not harming latency (\fig{fig:graph_batching_parameters}(b-c)). 
								 Notice how the optimal batching
								 time-window and model-allowed maximum batch size is determined
								 as a function of what the request traffic to the
								 inference server is, what the throughput-vs-latency tradeoff
								 curve is for a given processor architecture (\fig{fig:batching_impact}), and others.

\begin{figure}[t!] \centering
\includegraphics[width=0.485\textwidth]{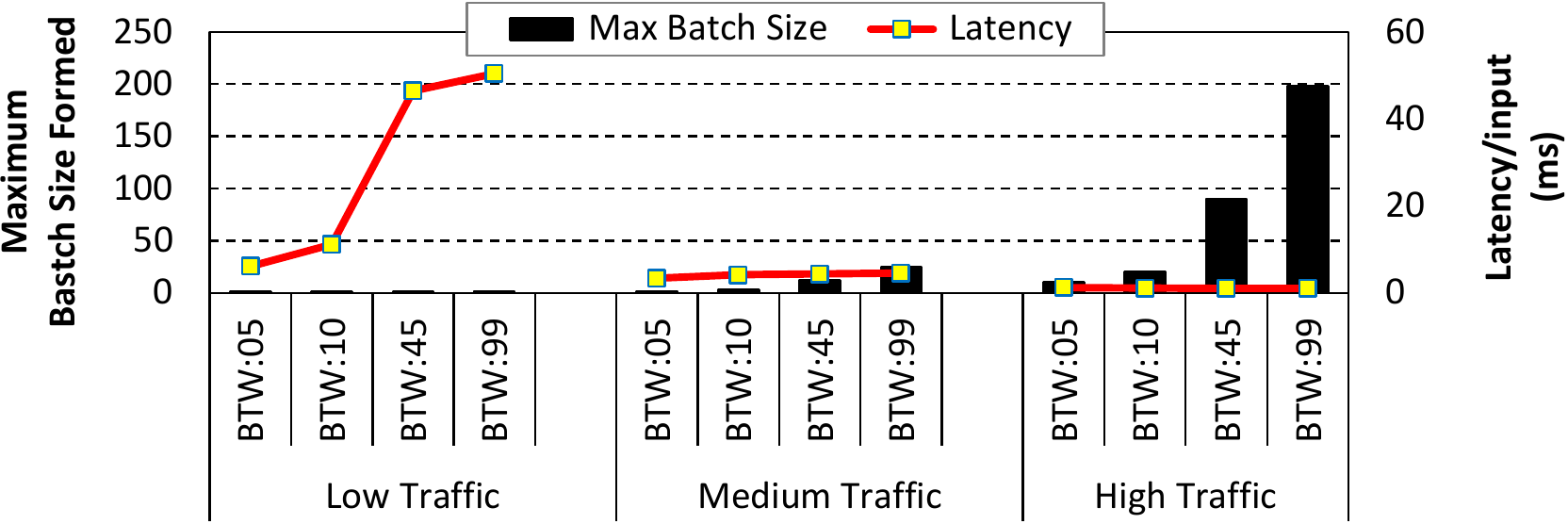}
\vspace{-0.5em}
\caption{
Effect of batching time-window (BTW, from $5$ ms to $99$ ms) on baseline graph batching's maximally formed
	batched size (left-axis) and average latency per input (right-axis) for ResNet, as a function of input request
	traffic load (x-axis). For low traffic conditions, a larger batching time-window
	does not help improve throughput and only end up harming average latency per input. Under heavy traffic,
	batching inputs starts being effective in improving throughput while still help reduce average latency
		per each input. This figure assumes $16$/$250$/$2000$ requests/sec query-arrival rate to model low/medium/high
		traffic. \sect{sect:methodology} details our methodology.
}
\vspace{-1.3em}
\label{fig:batch_sz_per_traffic}
\end{figure}

Overall, a fundamental challenge of graph batching is that the
\emph{statically} chosen batching time-window and maximum batch size is
utilized to handle \emph{all} scenarios, even though it is
practically impossible to estimate when the candidate inputs for batching will
arrive at the ML inference server.  When the server is lightly loaded, 
			 it is better to optimize the inference server for latency with
a short batching time-window with low maximum batch size. Conversely under
heavy request traffic, optimizing the inference server for both latency and throughput 
with a large enough batching time-window is preferrable (\fig{fig:batch_sz_per_traffic}). Unfortunately, the baseline
(static) graph batching by design cannot adapt to the dynamic request traffic
patterns.  Consider a scenario where the scheduler
just issued a new batched input prematurely (i.e., increasing batch size
		further helps improve throughput while having minimal impact on latency)
for execution because the batching time-window elapsed (\fig{fig:graph_batching_parameters}(b)).  If the server receives
new inputs just after such batched input was scheduled, a better scheduling
decision would have been to wait a bit longer (i.e., larger batching
		time-window) and seek a larger batch size
(\fig{fig:graph_batching_parameters}(c)). The static, ``one-size-fits-all''
approach of graph batching however is not able to properly handle the
aforementioned scenarios and reduce batching opportunities.

\subsection{Pitfalls of ``Application-Specific'' Batching}
\label{sect:limits_cellular_batching}

\begin{figure}[t!] \centering
\includegraphics[width=0.485\textwidth]{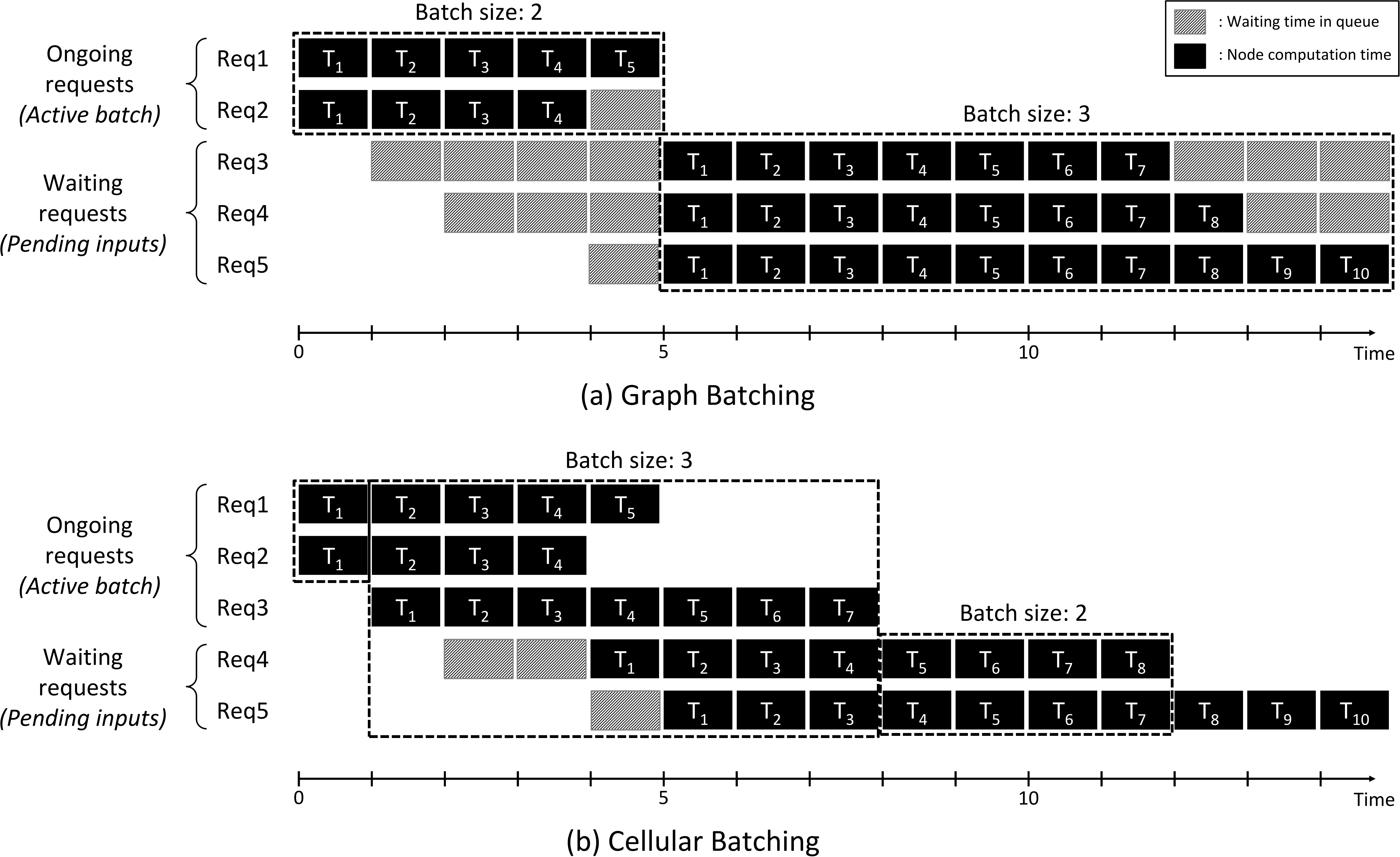}
\vspace{-0.5em}
\caption{
Timeline of  (a) graph batching and (b) cellular batching when three requests ($Req3$ to $Req5$)
	are received while $Req1$-$2$ are being processed. Figure assumes that each request
	is executing an RNN, each with a different output sequence length (determined by the number of times
			the recurrent layer is time-unrolled, e.g., $Req1$ with $5$ timesteps while $Req5$ with $10$ timesteps).
	The model-allowed maximum batch size is assumed to be configured to $3$, 
			which delays $Req4$ from being batched until t=4.
}
\vspace{-0.8em}
\label{fig:cellular_batching}
\end{figure}

\begin{figure}[t!] \centering
\includegraphics[width=0.485\textwidth]{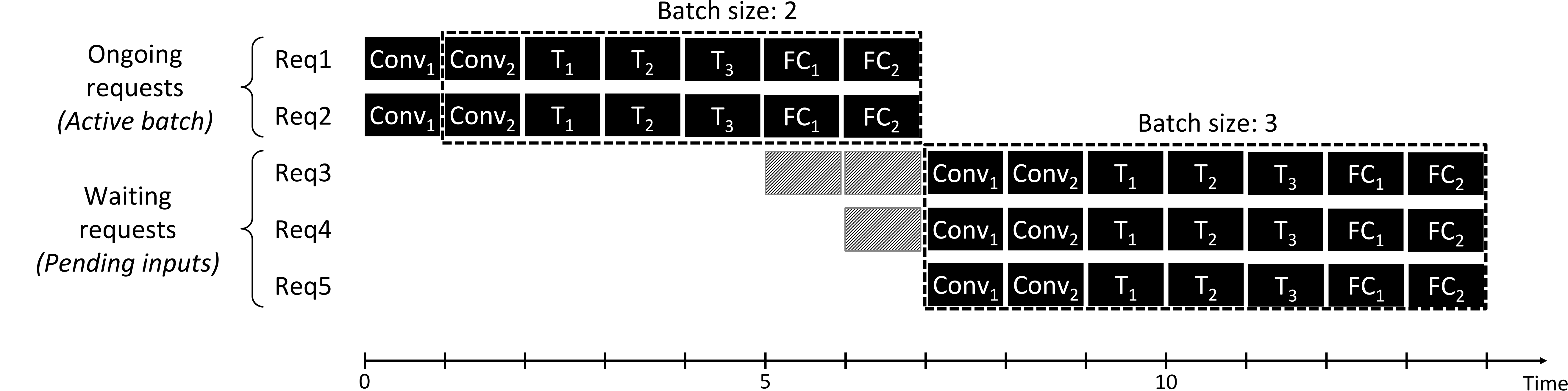}
\vspace{-0.3em}
\caption{
An example scenario where cellular batching fails to batch inputs (e.g., DeepSpeech-2). DNN graph
is assumed to have two convolutional (\texttt{CONV$_{i}$}) and two fully-connected layers (\texttt{FC$_{i}$})
before/after the recurrent layer.
}
\vspace{-1.5em}
\label{fig:cellular_failure}
\end{figure}

To tackle the limitations of graph batching, recent work by Gao et
al.~\cite{cellular_batching} proposed \emph{cellular batching}, which partially
addresses the batching problem from an application-specific perspective, with
an emphasis on RNN inference. A distinguishing feature of RNNs is that the RNN
cells within the time-unrolled recurrent layers all share the same weight values
across different timesteps (\fig{fig:cellular_batching}).  Cellular batching
utilizes such property to batch at the level of RNN cells rather than the
entire DAG, allowing new input requests to be batched into an ongoing batched
request. \fig{fig:cellular_batching} shows the different batching behavior
between graph vs. cellular batching. The baseline graph batching assumes that
the first $2$ requests ($Req1$-$2$) form a batch and starts execution at the
beginning of time. As the initial batched execution does not get completed
until t=$5$, the newly arrived requests ($Req3$-$5$) remain idle inside the
server, waiting for the current batch to finish execution. Cellular batching
can immediately schedule $Req3$ for execution as it can be batched with
$Req1$-$2$ at t=$1$ (and similarly, $Req4$ at t=$4$).  This is possible because
the unrolled RNN cells all share the same weight parameters across
different timesteps (e.g., $Req3$-$4$ and $Req5$ all execute using the same
		weights at t=$5$), enabling the batching system to more flexibly merge
requests at a fine-grained cell level. Overall, the benefit of cellular
batching is as follows.  First, 	cellular batching can reduce average response
time as the newly arrived requests can immediately join ongoing batched
requests without having to wait during the batching time-window.  Second, it
also helps improve system throughput as the likelihood of batching is
significantly improved thanks to the fine-grained, cell-level batching.
However, a key challenge of cellular batching is its limited applicability
among generic DNN workloads. Because cellular batching is specifically designed
to leverage the unique feature of RNNs (i.e., unrolled recurrent cells share
		the same parameters), the weight sharing effect it takes advantage of is no
longer applicable 	when the end-to-end DNN application contains \emph{non}-RNN
layers (e.g., convolutional or fully-connected layers). Consider the example
shown in \fig{fig:cellular_failure} which provides a high-level overview of
DeepSpeech-2's execution timeline using cellular batching. 
Once the first batch $Req1$-$2$ starts execution, cellular
batching is not able to batch the newly requested inputs $Req3$-$5$ into the ongoing
batch. This is because the future inputs $Req3$-$5$ must start executing from the first convolutional
layer yet the ongoing batch is already further down the execution process. As such,
		 cellular batching effectively levels down into the baseline graph
		 batching, \emph{serializing} the scheduling of $Req1$-$2$ and $Req3$-$5$ for DNNs
		 containing non-RNN layers within. 	

\subsection{Our Goal: A Flexible and Robust Batching System for ML}
\label{sect:cellular}

Overall, we observe several challenges with prior batching architectures.
First, baseline graph batching applies a brute-force, static solution to all deployment
scenarios which is suboptimal in handling the dynamic inference request traffic
patterns. Second, an application-specific batching solution like cellular
batching is optimized for a unique property of a specific (RNN) layer, so it can be
inapplicable for newly developed DNN layers or complex topologies
(\fig{fig:cellular_failure}).  Given how fast evolving the ML algorithmic
research space has been recently (e.g., state-of-the-art ML algorithms for NLP
		are no longer powered by RNNs but rather designed using attention
		modules~\cite{bert,gpt2}), a batching system tailored for a subset of the
DNN algorithms is unlikely to remain effective for the constantly evolving ML
research space. Lastly, it is of vital importance for end-users purchasing MLaaS to minimize 
SLA violations while maximizing
throughput for cost-efficiency. As we demonstrate in \sect{sect:results},
our SLA-aware batching can
	seamlessly adapt to the dynamic traffic patterns, 
achieving low latency while improving
	system throughput at all times. 
	Such property not only helps hyperscalars seeking to optimize TCO but also
	the end users of MLaaS. This is because low-latency and high-throughput can be achieved
	simultaneously, without having to painstakingly 
			fine-tune	batching time-window, maximum batch size, or other design parameters of
		graph batching, causing less burden to MLaaS consumers.

Our goal is to develop a batching system that can
flexibly adapt to the dynamically changing inference request traffic while
also being widely applicable for both current and future DNN topologies.  In
the following section, we detail our proposed batching architecture which
fundamentally addresses the limitations of prior batching solutions.

%% file: tex/proposed.tex
\section{LazyBatching: SLA-Aware Batching System for Cloud Machine Learning Inference}
\label{sect:lazyb_arch}

We propose \lazyb, an intelligent batching system that can dynamically adapt
its batching granularity to balance latency, throughput, and SLA satisfaction.

\begin{figure}[t!] \centering
\includegraphics[width=0.485\textwidth]{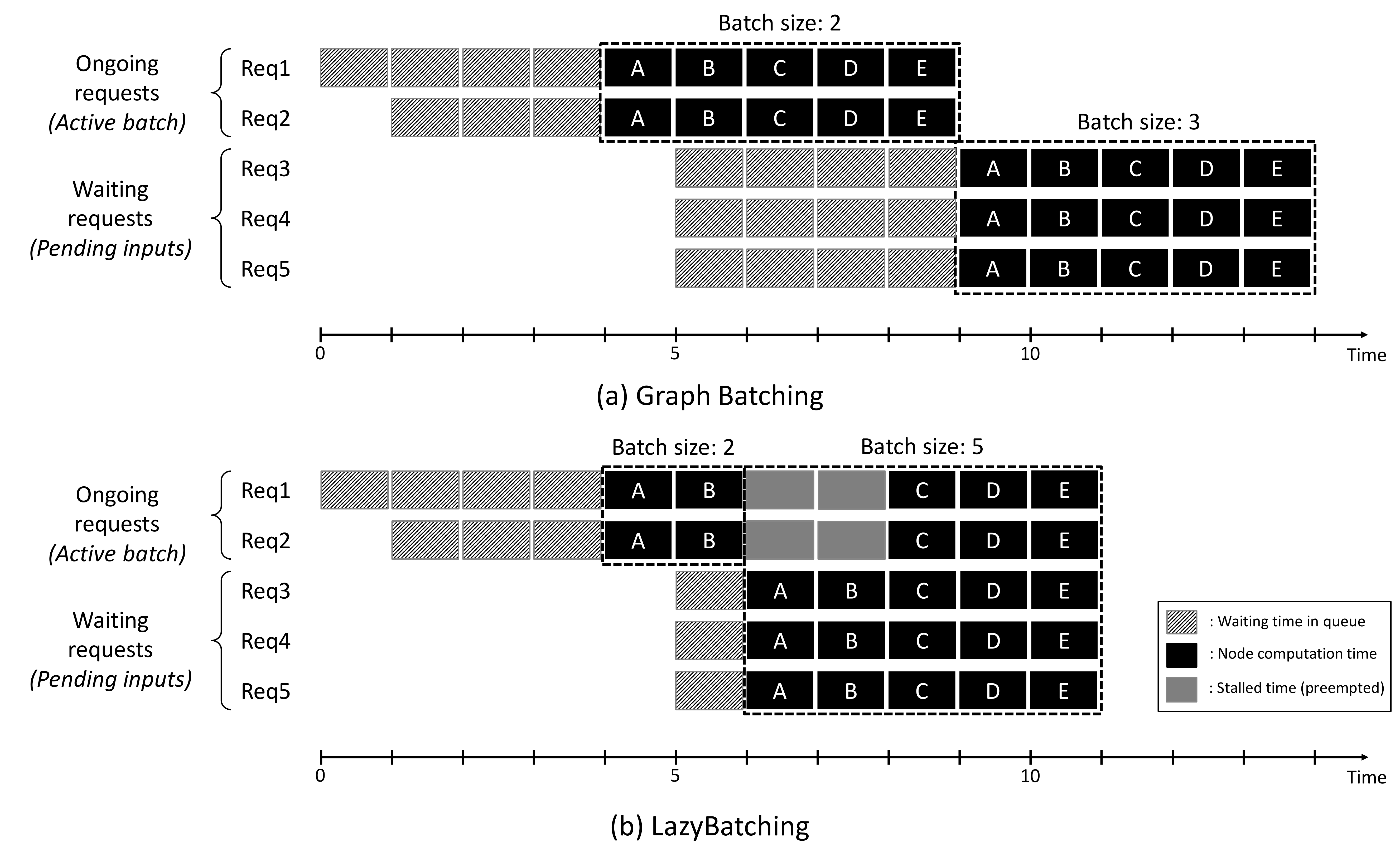}
\vspace{-0.5em}
\caption{
Proposed \lazyb execution timeline (vs. baseline). 
Each DNN is assumed to contain a fixed size of five graph nodes (node $A$ to $E$). 
}
\vspace{-1.3em}
\label{fig:lazyb_example}
\end{figure}

\subsection{Proposed Approach}
\label{sect:approach}

While the end-to-end DNN application  is represented as a graph structure, the
execution itself is conducted in a fine-grained \emph{node} (or layer)
granularity by the backend processor. Concretely, the runtime system in a
typical ML framework (e.g., TensorFlow, PyTorch, Caffe2) determines the
sequential order the graph nodes are to be executed for a target DNN model, and
schedules each individual nodes to the processing unit for execution
(\fig{fig:dnn_and_dag}).  As a result, user-level runtime APIs in popular backend DNN libraries such as
NVIDIA's cuDNN~\cite{cudnn} are designed in accordance to such node-level
execution model (e.g., \texttt{cudnnConvolutionForward()}).
Conventional batching systems however are based on a coarse-grained, \emph{graph-wide} scheduling
framework, which is at odds with the \emph{node-level} DNN execution model.
As highlighted in previous sections, a key limitation of graph batching comes
from its rigid, static graph-wide scheduling. Concretely, once a
batched graph is scheduled for execution, future inputs cannot execute until
the currently ongoing batch is finished. Such constraint  poses a fundamental
challenge in the batching opportunities between an ongoing batch and a newly
requested input because they cannot share a common layer (i.e., graph node) to
execute simultaneously.

	\begin{figure}[t!] \centering
\includegraphics[width=0.46\textwidth]{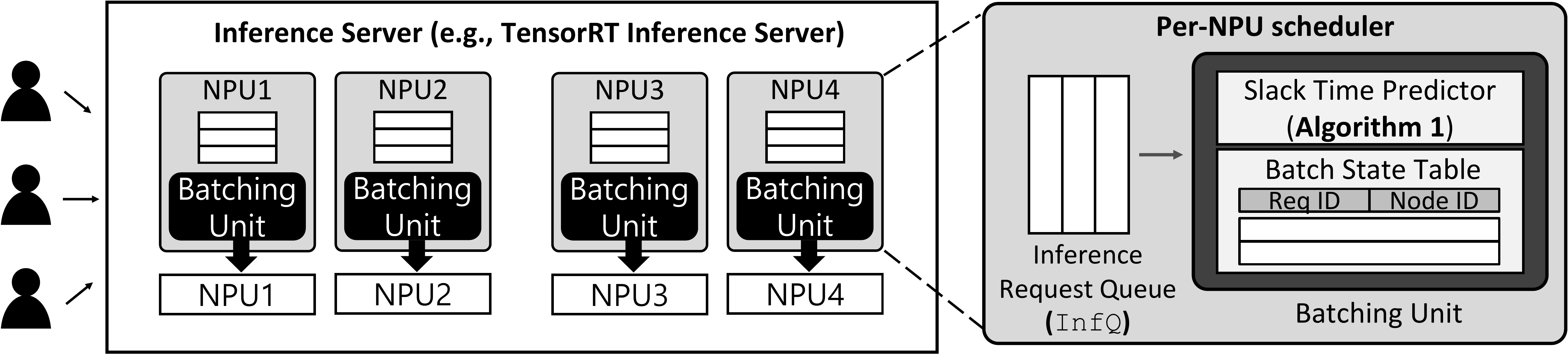}
\caption{
High-level overview of \lazyb model serving system.
}
\vspace{-1.3em}
\label{fig:lazyb_server_arch}
\end{figure}

	Rather than having a single batched input exclusively execute until
	completion, our {\bf key approach} is to maintain the scheduling granularity in a
	fine-grained \emph{node-level} and allow different (batched) inputs to be
	\emph{interleaved} for execution. \lazyb utilizes the node-level scheduling
	framework to preferentially schedule newly requested inputs to \emph{catch
		up} the progress of a previously ongoing, but yet to be finished earlier
		inputs. This opens up more batching opportunities as inputs can be 
		\emph{``lazily''} batched with each other in an incremental manner.  
		The notion of
		batching time-window is therefore non-existent with \lazyb because there
		is no fixed-length time window which inputs must wait in order to be
		batched together. In effect, our \lazyb scheduler constantly
		fires off one of the nodes within the pool of schedulable inputs, 
		whenever the batching  unit finds that appropriate to
		meet latency, throughput, and SLA goals.  \fig{fig:lazyb_example}
		illustrates an example where our scheduler virtually \emph{preempts}
		the execution of batched inputs $Req1$-$2$ at t=$6$ and \emph{context switch} to the
		execution of $Req3$-$5$ until it catches up the progress $Req1$-$2$ has made before it was
		preempted.  Once the context switched $Req3$-$5$ executes up to node $B$ at t=$8$,
		both $Req1$-$2$ and $Req3$-$5$ now share a common layer which our scheduler can
		safely \emph{merge} as a single batch to resume execution starting with node $C$.
			Note that the preemption of an ongoing batch, followed by a context switch to another (potentially batched)
	input, is \emph{always conducted in layer boundaries} by the runtime system at user-level because
	\lazyb naturally exploits the node-level scheduling framework
	(i.e., an ongoing batch will never
			get interrupted until its \emph{intra-node} computations are finalized). 	In other words,
	the node-level preemption and context-switching \emph{does not require any hardware modifications} 
	and is \emph{done purely in software} using existing ML frameworks and runtimes (\sect{sect:overhead} details the implementation overhead).

		Nonetheless, the effectiveness of lazily batching requests
		is dependent on whether the fine-grained interleaving of different input
		requests does not harm the responsiveness of individual inputs or violate
		SLA goals. A {\bf key innovation} of \lazyb is the development of an
		\emph{SLA-aware, slack time prediction model} which our scheduler utilizes
		to intelligently judge when/which inputs are worth lazily batching.
		In the rest of this section, we first detail our model serving 
		architecture followed by our slack time
		prediction model.

\subsection{LazyBatching Model Serving Architecture}
\label{sect:server_arch}

\fig{fig:lazyb_server_arch} provides a high-level overview of \lazyb's model
	serving system.  When the ML inference server receives an inference request,
	it is first forwarded to the inference request queue (\infq) and waits until
	the scheduler issues it (either in isolation or as a batch, grouped with other
			inputs) to the backend processor. 
	There are two key components that
	constitute our \lazyb server system.  First, our batching system maintains a
	\emph{batch state table} (\batchT) that tracks the batching status among the
	inputs currently executing.
	Second, an
	\emph{SLA-aware slack time predictor} is employed which utilizes
	domain-specific properties of ML inference to analyze whether lazily batching
	the currently executing inputs  and the ones waiting in the \infq (henceforth
			referred to as \emph{active batch} and \emph{pending
			inputs}, respectively) will result in an SLA violation or not.

When the SLA-aware slack time predictor determines that an additional 
	batching  can violate SLA, then the scheduler does not try to batch
	more inputs and authorize the currently active batch to complete its
	execution uninterrupted.  However, if the likelihood of an
	SLA violation through lazily batching is low, then our scheduler
	first preempts the active batch at the end of the current node.
	It then context switches to the pending inputs to allow it to catch up the
	progress of the preempted, previously active batch.  During the course of
	this process, the \batchT keeps track of the layer-wise execution status of
	the preempted and preempting inputs so that they can be batched together once
	they reach the same graph node.  Below we first discuss how \lazyb utilizes
	the \batchT for node-level scheduling and batching.

	\begin{figure}[t!] 
\centering
\subfloat[]{
\includegraphics[width=0.485\textwidth]{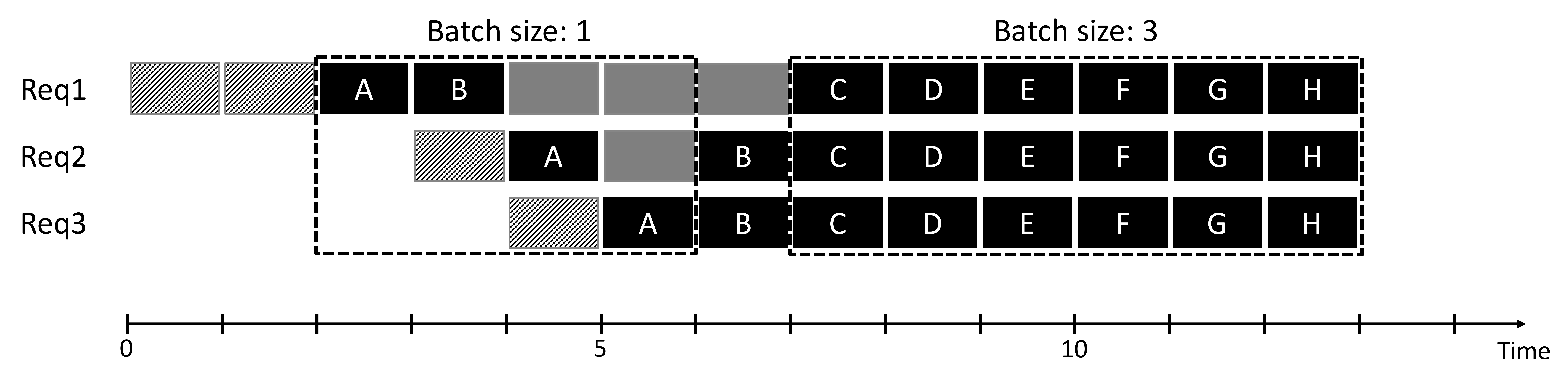}
	\label{fig:lazyb_xx}
}
\vspace{-0.3em}
\subfloat[]{
	\includegraphics[width=0.485\textwidth]{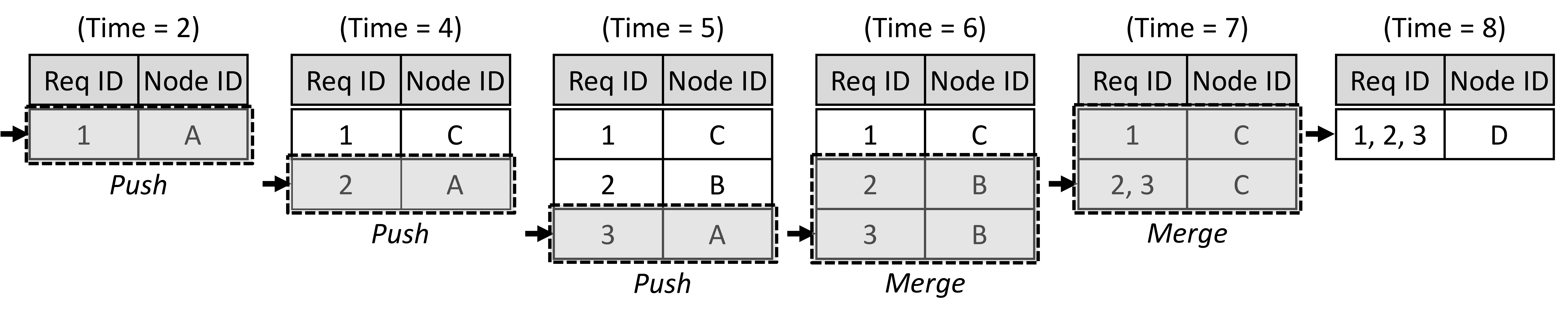}
	\label{fig:stack_example}
}
\vspace{-0.3em} 
\caption{ 
(a) Timeline of \lazyb when executing a graph with $8$ nodes, and
	(b) the changes in \batchT as stack entries are \emph{pushed}/\emph{merged}.
Example assumes that the slack time predictor always 
	find lazily batching pending input requests ($Req2$-$3$) with the
	active batch ($Req1$) beneficial.
The stack grows from top to bottom in this figure (i.e., the black arrow points to the top of the stack).
}   
\vspace{-1.55em}
\label{fig:lazyb_impact}
\end{figure}

{\bf Stack-based batch status tracking.} \lazyb implements
	the \batchT as a software \emph{stack} data structure and the
	entry at the top of the stack corresponds to the active batch that is
	currently executing. 	Each stack entry tracks
	what is the graph node ID a group of batched inputs
	(referred to as \emph{sub-batch}) will be executing. \lazyb 
	utilizes the \batchT to examine a sub-batch's basic requirements to be batched with
	other sub-batches (i.e., whether they are able to execute a common node).  \fig{fig:lazyb_impact} shows how the  \batchT is utilized
	to lazily batch incoming requests on-the-fly.  Because
	the \infq only has a single input $Req1$ initially, the top of the
	stack entry (one corresponding to this sub-batch) is \emph{pushed} with a
	request ID of $1$ and node ID $A$ as shown in t=2 of
	\fig{fig:lazyb_impact}(b). Suppose the inference server receives another
	input $Req2$ while $Req1$ is busy executing node $B$, and the SLA slack 
	predictor deems it advantageous to merge $Req1$ and $Req2$ as a single batch.
	In situations like this, our scheduler first updates the next graph node ID
	of our active batch $Req1$ to node $C$ at the end of node $B$'s execution to
designate the fact that this sub-batch should execute node $C$ once the
scheduler issues it again to the processor.  The scheduler then preempts the
execution of $Req1$ and pushes another
stack entry corresponding to $Req2$ (i.e., request ID of $2$ executing node
		$A$) to the \batchT so that $Req2$ becomes the \emph{new} active
batch to be issued to the processor (at t=4 of \fig{fig:lazyb_impact}(b)).
As the scheduler context switches to $Req2$ and executes node $A$, the server receives another request $Req3$,
			which the slack predictor decides to lazily batch it with
			$Req1$-$2$. This is done by again preempting $Req2$ at t=5 when it
			finishes executing node $A$, and then pushing another stack entry of
			$Req3$ to have the scheduler execute $Req3$ afterwards.  Once the new
			active batch $Req3$ finishes executing up to node $A$, the node ID field in
			the stack is updated to $B$ (t=6 at \fig{fig:lazyb_impact}(b)). Notice
			how the node ID field of the two topmost stack entries are now
			identically at node $B$, meaning all the inputs that are part of these
			two sub-batches can be merged as a single batch. The batching of these
			two entries is undertaken by \emph{merging} the two topmost stack entries
			as a single one, as illustrated in t=6 of \fig{fig:lazyb_impact}(b),
			which allows both $Req2$ and $Req3$ to execute concurrently starting
			graph node $B$. \fig{fig:lazyb_impact}(b) similarly shows the updates to
			the \batchT when the batched $Req2$-$3$ gets lazily batched again with
			$Req1$ at t=7.  
			Because the stack push/merge operations are 
			only invoked at layer-boundaries in software,
			\batchT enables a low-cost yet high performance
			control mechanism to track batching status.

\subsection{``SLA-Aware'' Slack Time Prediction}
\label{sect:slack_estimation}

Providing fast user responsiveness is of utmost importance for user-facing ML
inference, so cloud service providers typically have 
	SLA targets to meet to satisfy QoS requirements. \lazyb utilizes our \emph{SLA-aware slack time
		prediction} model to only authorize batching when it
		will not violate the SLA.  Our prediction model
		quantifies how much slack time a given batched input has remaining before
		violating a model-specific SLA target.  The estimation of a batched input's
		slack time is done conservatively (i.e., predict that SLA slack time is
				smaller than what actually remains) such that the scheduler is optimized
		to minimize the number of SLA violations first and improve throughput second.
		Our SLA-aware slack time estimator consists of three key
		components: 1) node-level latency estimation, 2) graph-wide estimation, and
		3) utilizing these two components for slack estimation. We first detail our
		definition of SLA slack time, followed by a description of our
		node-level/graph-wide latency estimation model.

{\bf Slack time prediction.}	Consider the first request $Req1$ in
\fig{fig:lazyb_impact}, which we use as a running example to explain our slack
model.  If the processor is currently busy handling other requests,
					 $Req1$ will have to wait in \infq until it gets issued to
					 the processor for the first time (two time-units, from t=$0$ to $2$
							 in \fig{fig:lazyb_impact}(a)).  Because the initial server wait
					 time ($T_{wait}$) also counts against SLA, our model needs to
					 subtract $T_{wait}$ from the model-specific, constant SLA value
					 ($SLA_{target}$) when estimating slack.  Once $Req1$ starts
					 execution, $Req1$'s remaining slack becomes a function of how long
					 it will take for $Req1$ to complete the end-to-end DNN execution.
					 Accordingly, the slack time of $Req1$ \emph{without} batching is:

\vspace{-1em}
\begin{equation}
\label{eqn:slack_alone}
\footnotesize
Slack=SLA_{target}-(T_{wait}+SingleInputExecTime_{Req1}) 
\end{equation}
\vspace{-1em}

\begin{algorithm}[t!]
\caption{DNN graph-wide inference time estimation}
\label{algo:graph_latency}
\begin{algorithmic}[1]
\footnotesize
\STATE $SingleInputExecTime \gets 0$
\STATE $GraphLatency \gets 0$
\FOR {$n \ \textbf{in} \ nodes$}
\IF {$Type(n) \ \textbf{is} \ STATIC$}
\STATE $GraphLatency \ += NodeLatency(n)$
\ELSIF {$Type(n) \ \textbf{is} \ ENCODER$}
\STATE $GraphLatency \ += NodeLatency(n) \times enc\_timesteps$
\ELSE
\STATE $GraphLatency \ += NodeLatency(n) \times dec\_timesteps$
\ENDIF
\ENDFOR
\STATE $SingleInputExecTime \gets GraphLatency$
\STATE \RETURN {$SingleInputExecTime$}
\end{algorithmic}
\end{algorithm}

For an $SLA_{target}$ of $30$ time-units, then the slack time without batching is estimated as
``$30$$-$($2$$+$$8$)$=$$20$'' for the given examples in \fig{fig:lazyb_impact}
(i.e., $8$ time-units is consumed when $Req1$ executes node $A$ to $H$). However, under
a scenario where $Req1$ \emph{is} batched with $Req2$, 
the \emph{SingleInputExecTime} term in \eqn{eqn:slack_alone} should incorporate
the batching effects for slack estimation.  If we were to have the exact
throughput-vs-latency tradeoff curves for every graph node within the target
DNN model (similar to \fig{fig:batching_impact}, but evaluated for every graph
		node under all possible inference batch size), we can quantify the impact
the potential (lazy) batching between an active batch ($Req1$) and pending
inputs ($Req2$) will have on end-to-end inference latency. Maintaining such
\emph{oracular} tradeoff curve for all possible graph nodes and batch size
however is cumbersome let alone requires a high implementation overhead.
As the primary goal of our slack estimation is to minimize SLA violations, we propose
to \emph{conservatively} estimate the inference latency of \emph{batched} inputs
as a summation of all input's \emph{single-batch latency, executed in isolation}.
While this \emph{overprovisions} the estimated inference time of \emph{batched} inputs, it helps reduce 
the estimated slack time thereby reducing the likelihood of SLA violations.
\eqn{eqn:inf_latency} summarizes our slack time prediction model which assume
the initial input ($Req1$) is batched with $(N-1)$ future requests: 

\vspace{-1em}
\begin{equation}
\label{eqn:inf_latency}
\footnotesize
Slack=SLA_{target}-(T_{wait}+\sum_{i=1}^{N}{SingleInputExecTime_{i}})
\end{equation}
\vspace{-0.5em}

In \sect{sect:results}, we quantitatively demonstrate that our conservative
slack estimation model is competitive even compared to its
\emph{oracular} version, which utilizes the aforementioned oracular tradeoff
curve in estimating a batch's precise execution time.  As both $SLA_{target}$
and $T_{wait}$ are known values, deriving the $Slack$ value in
\eqn{eqn:inf_latency} requires an estimation of an individual, single-batched
input's end-to-end, graph-wide execution time (i.e.,
		$SingleInputExecTime_{i}$). We now discuss our node-level/graph-wide
latency estimation model for predicting $SingleInputExecTime_{i}$ (\algo{algo:graph_latency}).

{\bf Node-level latency estimation.} Our key observation is that each
individual graph node's execution time over a target hardware architecture is
highly deterministic and predictable.  A graph node's layer configuration is
determined at compile time and the layer weight values are also statically
fixed for inference. As a result, the computation and memory access
characteristics of a graph node (i.e., DNN layer) is highly regular and
input-independent, exhibiting little per-layer latency variation across
different executions. Prior work~\cite{jointdnn,jia2018beyond} similarly observed the
deterministic nature of DNN inference and our node-level latency
estimator exploits such property. We therefore propose to
\emph{profile} the per-node execution time of the target DNN and characterize its
average per-node latency as a software-level lookup table. The node-level
latency lookup table ($NodeLatency(n)$ in \algo{algo:graph_latency}) is then
utilized for estimating the DNN's graph-wide execution time.  The profiling overhead
is negligible as the characterization of a
DNN graph's node-level latency only has to be done once and be reused for all future inferences
for that model.

{\bf Graph-wide latency estimation.}	Predicting the graph-wide execution time
requires an estimation of how \emph{many} graph nodes to traverse for a given DNN's
inference. As discussed in \sect{sect:dnn}, DNNs with a \emph{static}
graph topology have a fixed number of nodes to execute,
			irrespective of what the input value is. Consequently, estimating the
			graph-wide inference time of a static DNN (e.g., CNNs) is
			straighforward where we simply conduct a summation of all the node-level
			latency estimations as summarized in line $3$-$5$ of \algo{algo:graph_latency}.

				\begin{figure}[t!] \centering
		\includegraphics[width=0.485\textwidth]{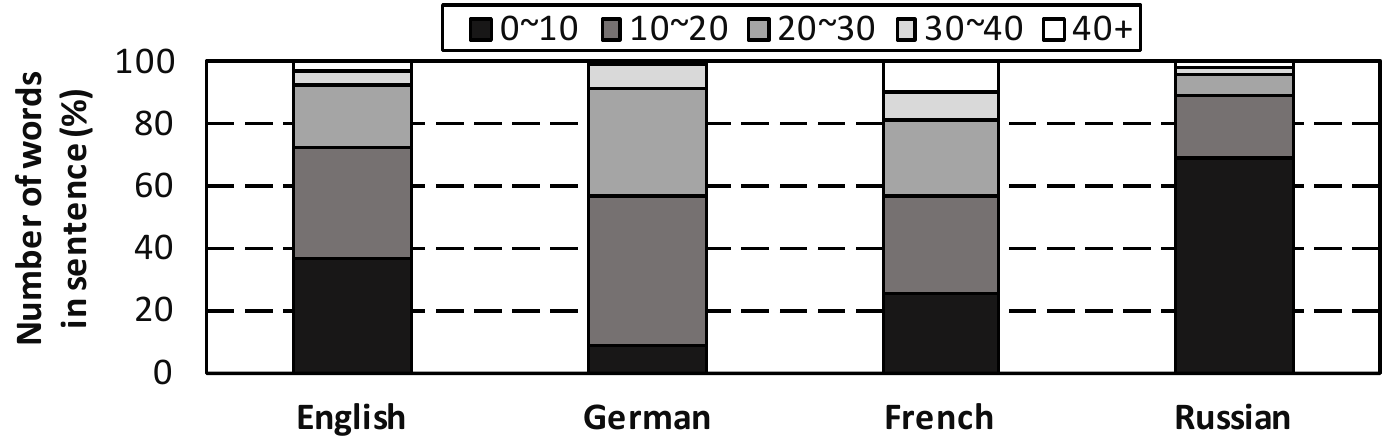}
		\vspace{-0.5em}
		\caption{
		Number of words within a sentence when characterized across 	
		WMT-2019's $30,000$ ``English-to-German/French/Russian'' translation pairs~\cite{wmt}.
		}
		\vspace{-1.65em}
		\label{fig:profiling}
		\end{figure}

			However, precisely estimating the latency of a \emph{dynamic} graph DNN
			is challenging, if not impossible, because the number of nodes to
			traverse within the DAG is variable and input-dependent. Consider the
			English-to-German translation example shown in \fig{fig:dag}.  Depending
			on what the input (English) sequence length is, the output (German)
	sequence length can vary, represented by the number of times the
	recurrent layer in RNNs (or the decoder block in attention
			modules~\cite{transformer,bert}) have been time-unrolled. Because the
	number of unrolled timesteps (i.e., the number of translated German words) is
	determined dynamically at runtime, statically estimating the graph-wide
	inference time is challenging.  
	
	Nonetheless, recall that our primary
	scheduling objective is to minimize SLA violations and our slack time
	prediction model is devised in accordance with that principle (i.e.,
			overestimate a batched input's execution time for a conservative slack
			prediction, \eqn{eqn:inf_latency}).
	As such, we propose a profile-driven characterization based approach that
	sufficiently overprovisions the dynamic DNN's graph-wide estimated latency to
	minimize the likelihood of SLA violations.  The key intuition behind our
	proposal is that the number of times the dynamic DNN will be unrolled into
	(i.e., the output sequence length in language translation examples) is
	determined by \emph{how} the DNN model has been trained.  As the training
	dataset determines how the model gets trained (and accordingly the model's
			inference time behavior), a detailed characterization across the training
	dataset can provide a statistical guideline on what is the likelihood of the
	trained model's recursive layer to be unrolled into a particular output
	sequence length (i.e., the unrolled decoded sequence number).  In other
	words, the time-unrolled recurrence length  will likely fall within the set
	of output sequence lengths that we observed during the training dataset
	characterization.  \fig{fig:profiling} summarizes the result of our
	characterization study which shows what fraction of the
	training dataset contains sentences with a particular output sequence length.
	For example, approximately $70\%$ of the English sentences in WMT-$2019$ training
	 dataset have less than $20$ words.  Using such profiled	information, our proposed
	 approach is to \emph{statically} choose a maximum output sequence length value
	($dec\_timesteps$ in \algo{algo:graph_latency})
	that sufficiently covers more than $N$-$\%$ of the decoded, output sequence length
	as observed in the characterization study.
	For instance, approximately $90\%$ of the translated German sentence word
	count will likely fall within $30$ words, so having the $dec\_timesteps$
	value be statically set as $30$ words (i.e., assume $N$$=$$90\%$) will allow the
	scheduler to conservatively estimate the graph-wide latency (line $8$-$9$ in
			\algo{algo:graph_latency}). If the output sequence length were to be
	evaluated smaller than $dec\_timesteps$ at runtime (e.g., less than $10$
			words in the translated German sentence), then the $GraphLatency$ is
	overly estimated which eventually reduces the estimated slack time. Such
	conservative estimation of slack time however helps minimize SLA violations,
	which is our first and foremost scheduling objective.  The default configuration of
	\lazyb is to set $N$$=$$90\%$ but service providers can use the value
		of $N$, and accordingly the $dec\_timesteps$ value, as a tuning knob to balance
			SLA violations and throughput.  In \sect{sect:eval_sensitivity}, we
			quantitatively discuss the sensitivity of \lazyb to $dec\_timesteps$ and
			demonstrate the robustness of our prediction model.
	
\begin{table}[t!]
  \centering
  \caption{NPU simulator configuration.}
\scriptsize
\vspace{-0.5em}
  \begin{tabular}{|c|c|}
		\hline
		\multicolumn{2}{|c|}{\textbf{Processor architecture}} \\
		\hline
    Systolic-array dimension     			& $128 \times 128$   \\
    \hline              
    Operating frequency				& $700$ MHz \\
    \hline              
    On-chip SRAM size (activations \& weights)   			& $8$ \& $4$ MB \\
    \hline              

    \multicolumn{2}{|c|}{\textbf{Memory subsystem}} \\
    \hline
    Number of memory channels 	& $8$	\\
    \hline
    Memory access latency & $100$ cycles  \\
    \hline
    Memory bandwidth	& $360$ GB/sec	\\
		\hline
  \end{tabular}
\vspace{-1.85em}
  \label{tab:npu_config}
\end{table}

\subsection{Putting Everything Together}
\label{sect:everything_together}

Overall, our SLA-aware slack time predictor (\eqn{eqn:inf_latency}) utilizes
domain-specific properties of ML inference (\algo{algo:graph_latency}) for
a conservative estimation of slack time, only authorizing batching when the likelihood
of an SLA violation is low.
 The \lazyb scheduler then utilizes the
software-level \batchT as a lightweight control mechanism
(\fig{fig:lazyb_impact}) for node-level scheduling and batching of
active/pending inputs. Compared to baseline graph batching, our proposal can flexibly
adapt the level of batching per server inference queries and achieve high system throughput
while significantly reducing the number of SLA violations. 

%% file: tex/methodology.tex
\section{Methodology}
\label{sect:methodology}

As discussed in \sect{sect:scope}, our study assumes NPUs as the baseline architecture
Due to the lack of publicly accessible NPUs, we resort to
a simulation based evaluation methodology in our default settings.
The applicability and effectiveness
of \lazyb over real GPU-based inference systems is quantitatively 
demonstrated later in \sect{sect:eval_sensitivity}).

{\bf Simulation methodology.} The baseline NPU architecture is modeled after Google's TPU
design, which employs a systolic-array based
microarchitecture~\cite{tpu1,tpu2}.  We designed our cycle-level performance
model based on \cite{tpu1} as well as public patents from
Google~\cite{tpu_patent1,tpu_patent2,tpu_patent3,tpu_patent4}. The performance
model has been cross-validated against Google Cloud TPU~\cite{cloud_tpu} and
SCALE-Sim~\cite{scalesim}, an open-sourced systolic-array based NPU
simulator.
Because the
		compute and memory access characteristics of DNNs  exhibit a deterministic
		dataflow with high data locality, the system-level performance is less
		sensitive to the complex behavior of the DRAM microarchitecture (e.g., row
				open/close, refresh, $\ldots$). Following prior
		work~\cite{scnn,cnvlutin,stripes}, we modeled the memory system as having
		fixed latency and memory bandwidth	to reduce simulation time (\tab{tab:npu_config}).

{\bf Benchmarks.} We employ the
		methodology employed in MLPerf cloud inference benchmark
		suite~\cite{mlperf} to generate inference
		request traces.  Concretely, we establish an inference query traffic
		generator which issues inference requests to the model serving system based
		on a \emph{Poisson distribution} to emulate a server's query-arrival rates
		as in other relevant prior work~\cite{cellular_batching,
			nexus, shenango, optimus}.  The parameters of our Poisson distribution
			are chosen to model the low/medium/heavy load traffic to the inference
			server (i.e., $0$-$256$/$256$-$500$/$500$+ queries/sec for
					low/medium/heavy traffic), in accordance with the single-input
			inference latency of our studied workload, which ranges from $1-7$ ms
			(\tab{tab:benchmarks}).  
			In terms of the evaluated benchmarks, the
				main evaluation section (\sect{sect:eval_latency} and
						\sect{sect:eval_sla}) primarily focuses on three workloads
					summarized in \tab{tab:benchmarks} for a detailed analysis of
					\lazyb's effectiveness across different dimensions.
					We select two applications from the MLPerf
					inference benchmark suite used for computer vision (ResNet) and
					machine translation (GNMT).  
	We also					study an attention-based machine translation application
					(Transformer) included as part of the MLPerf training benchmark
					suite, which we utilize for inference.  Both GNMT and Transformer
					assume an English-to-German sentence translation scenario with a
					maximum sentence length of 80 words. Later in \sect{sect:eval_sensitivity},
						we quantify the robustness 
						of \lazyb across a broader set of applications by studying its performance
						across four additional benchmarks (i.e.,
						VGGNet~\cite{vggnet}, MobileNet~\cite{mobilenet}, Listen-Attend-and-Spell~\cite{las},
						and BERT~\cite{bert}) during our sensitivity analysis.  To model the
						\emph{predicted} and \emph{actual} time-unrolled output sequence
						length of seq2seq models, we take the following measure. For a
						given single-input inference query, we randomly select an English
						sentence from the WMT-2019 test dataset (unused as part of the
								profile-based characterization study which uses the training
								dataset only). The selected English sentence is translated into
						its corresponding German sentence, which we utilize to count the
						number of its words and use it to model the \emph{actual}
						time-unrolled output sequence at runtime.  As discussed in
						\algo{algo:graph_latency}, the \emph{predicted} output sequence
						length (i.e., $dec\_timesteps$) is fixed at a static threshold
						value assuming $N$$=$$90\%$ coverage of our profile-driven
						characterization study (\fig{fig:profiling},
								\sect{sect:slack_estimation}).  The sensitivity of \lazyb to
						other translation pairs and alternative $dec\_timesteps$ values are
						discussed in \sect{sect:eval_sensitivity}. 
					
\begin{table}[t!]
  \centering
  \caption{Evaluated benchmarks.}
\scriptsize
\vspace{-0.5em}
  \begin{tabular}{|c|c|c|c|}
		\hline
    \textbf{Network name}  & \textbf{Application}  &  \textbf{ML algorithm} & \textbf{Single-batch latency} \\
		\hline
    ResNet~\cite{resnet}   & Vision  			&   CNN  & $1.1$ ms \\
    \hline              
    GNMT~\cite{gnmt}		   & Translation   		  & RNN & $7.2$ ms\\
    \hline              
    Transformer~\cite{transformer} & Translation 		 & Attentions & $2.4$ ms \\
    \hline              
  \end{tabular}
\vspace{-1.85em}
  \label{tab:benchmarks}
\end{table}

%% file: tex/result.tex
\section{Evaluation} 
\label{sect:results}

\begin{figure}[t!] 
\centering
\subfloat[ResNet]{
\includegraphics[width=0.485\textwidth]{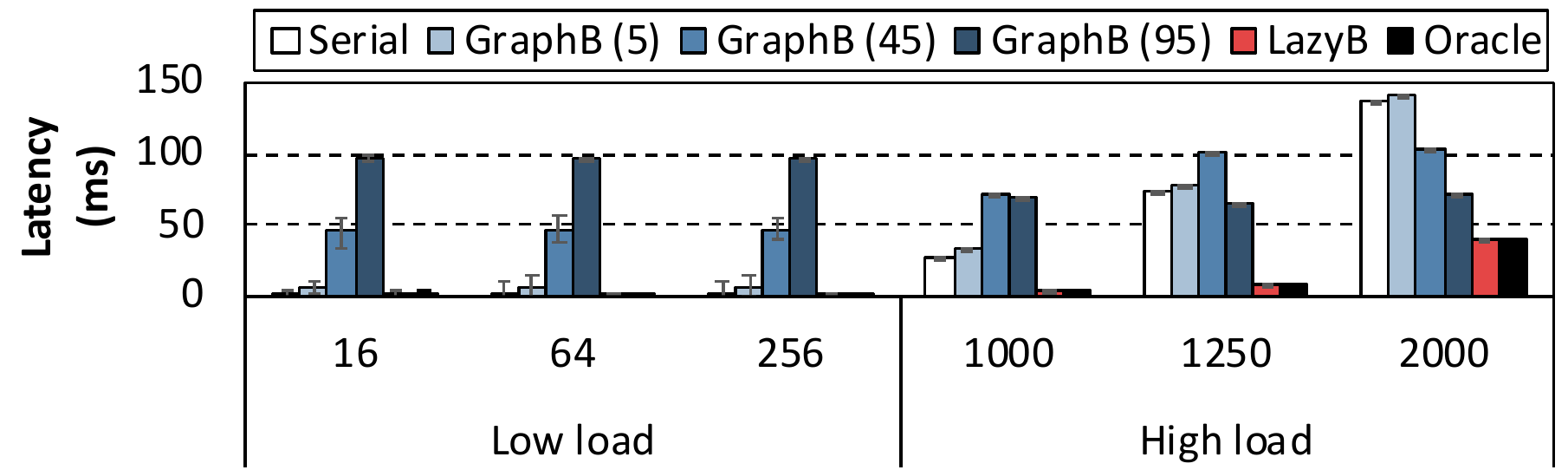}
	\label{fig:latency_resnet}
}
\vspace{-0.1em}
\subfloat[GNMT]{
	\includegraphics[width=0.485\textwidth]{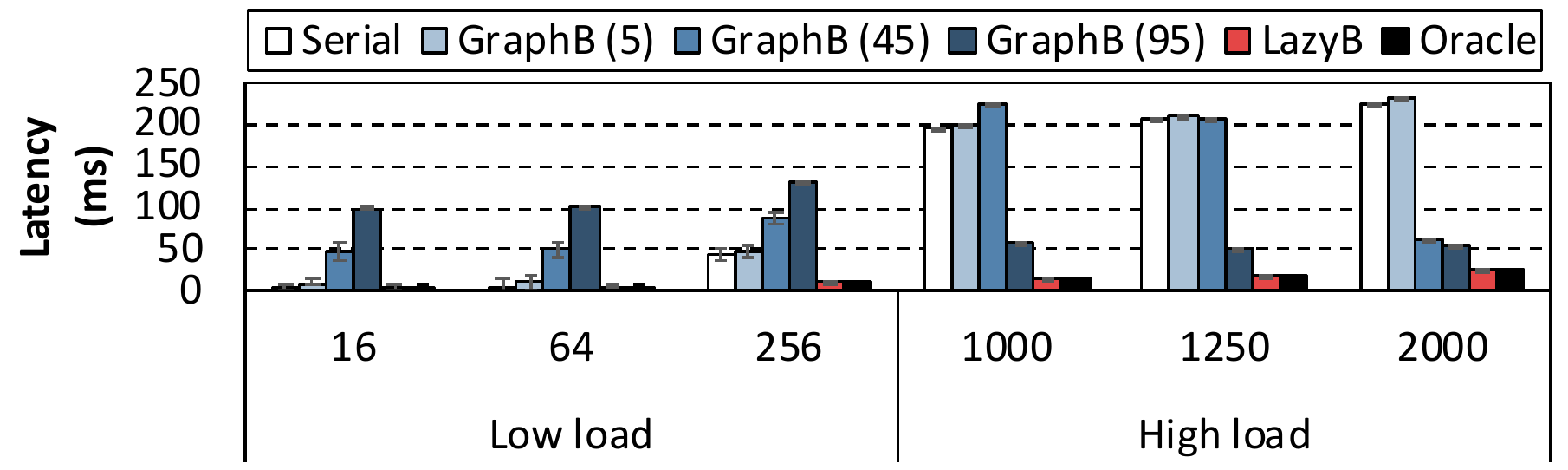}
	\label{fig:latency_gnmt}
}
\vspace{-0.1em}
\subfloat[Transformer]{
	\includegraphics[width=0.485\textwidth]{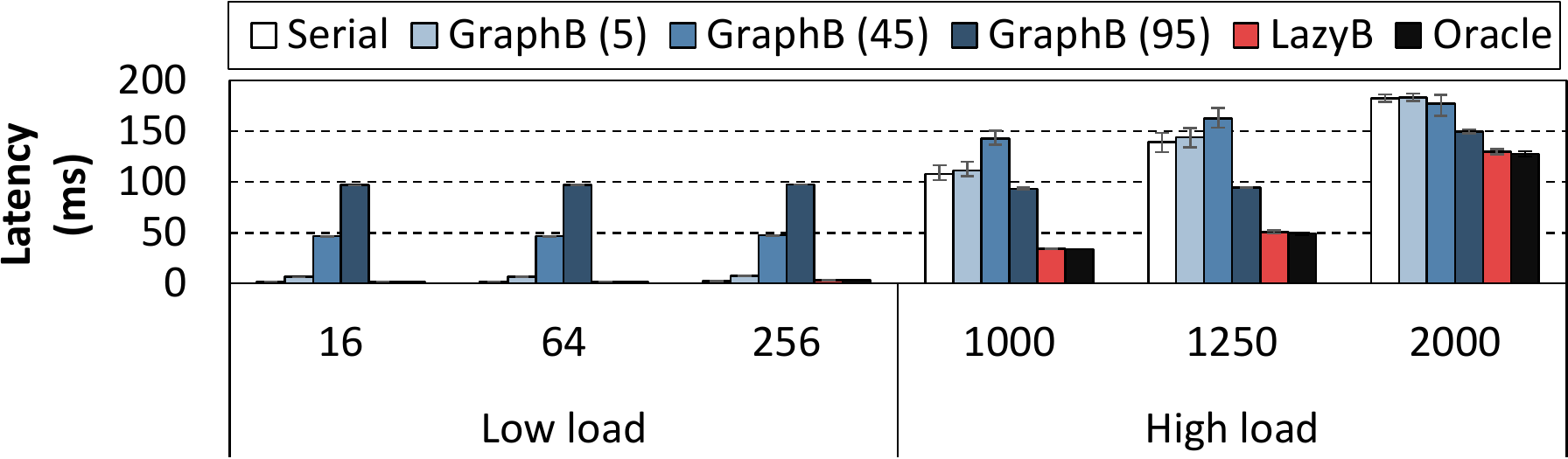}
	\label{fig:latency_transformer}
}
\vspace{-0.2em} 
\caption{ 
Effect on average latency per query-arrival rate (x-axis, requests/sec). 
}   
\vspace{-1.3em}
\label{fig:eval_latency}
\end{figure}

We explore four design points in this section: 1) always \emph{serializing}
incoming requests without batching (\texttt{Serial}), 2) baseline graph
batching with a batching time-window of $N$ ms (\texttt{GraphB(N)}), 3) our
proposed \lazyb (\texttt{LazyB}), and 4) an oracular version of \lazyb
(\texttt{Oracle}) that utilizes the precise latency-vs-throughput tradeoff
curves (for all possible batch sizes for every node within a target DNN) to
estimate SLA slack time and perform lazy batching. For clarity of explanation,
				 graph batching is configured with a model-allowed maximum batch size
				 of $64$ as default, but we discuss the sensitivity of our results
				 against other maximum batch sizes in \sect{sect:eval_sensitivity}.  As
				 SLA target numbers are vendor-specific values not publicly disclosed,
				 we assume the SLA deadline is set to $100$ ms for \lazyb's slack estimation 
				 in \sect{sect:eval_latency}.  The effectiveness of \lazyb
				 under different SLA targets is discussed in \sect{sect:eval_sla}.  We
				 omit the results of cellular batching because none of the workloads we
				 study are solely based on RNN layers, rendering \emph{cellular batching to
				 perform identically to graph batching}.  This section reports the
				 averaged results across $20$ simulation runs.


\subsection{Effect on Latency and Throughput}
\label{sect:eval_latency}

\begin{figure}[t!] 
\centering
\subfloat[ResNet]{
\includegraphics[width=0.485\textwidth]{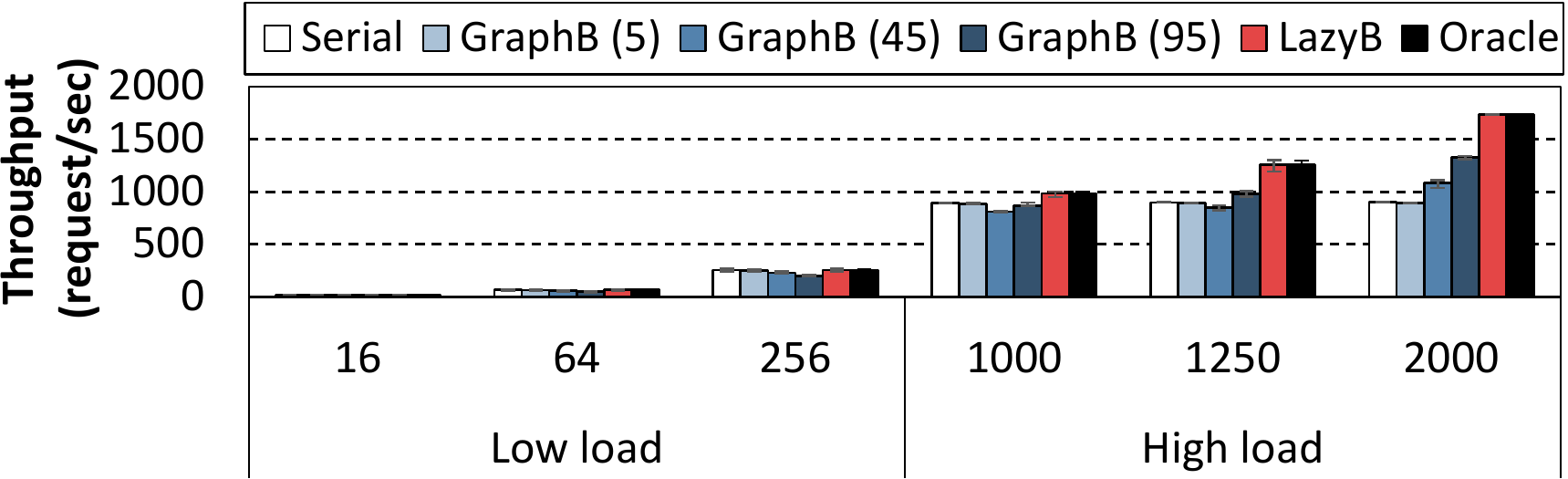}
	\label{fig:throughput_resnet}
}
\vspace{-0.1em}
\subfloat[GNMT]{
	\includegraphics[width=0.485\textwidth]{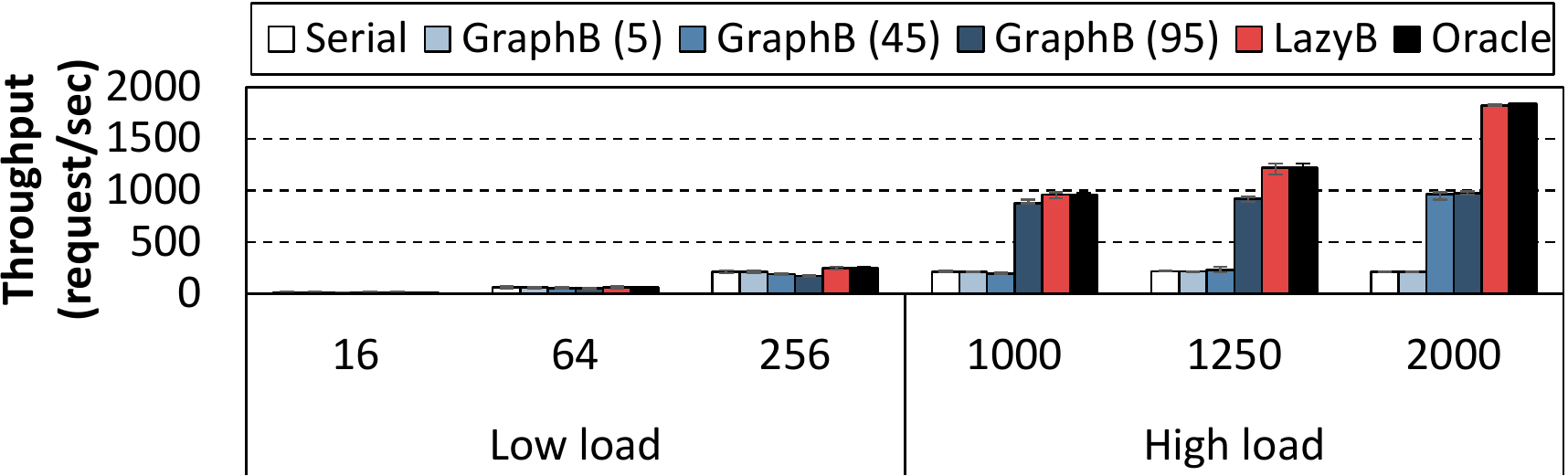}
	\label{fig:throughput_gnmt}
}
\vspace{-0.1em}
\subfloat[Transformer]{
	\includegraphics[width=0.485\textwidth]{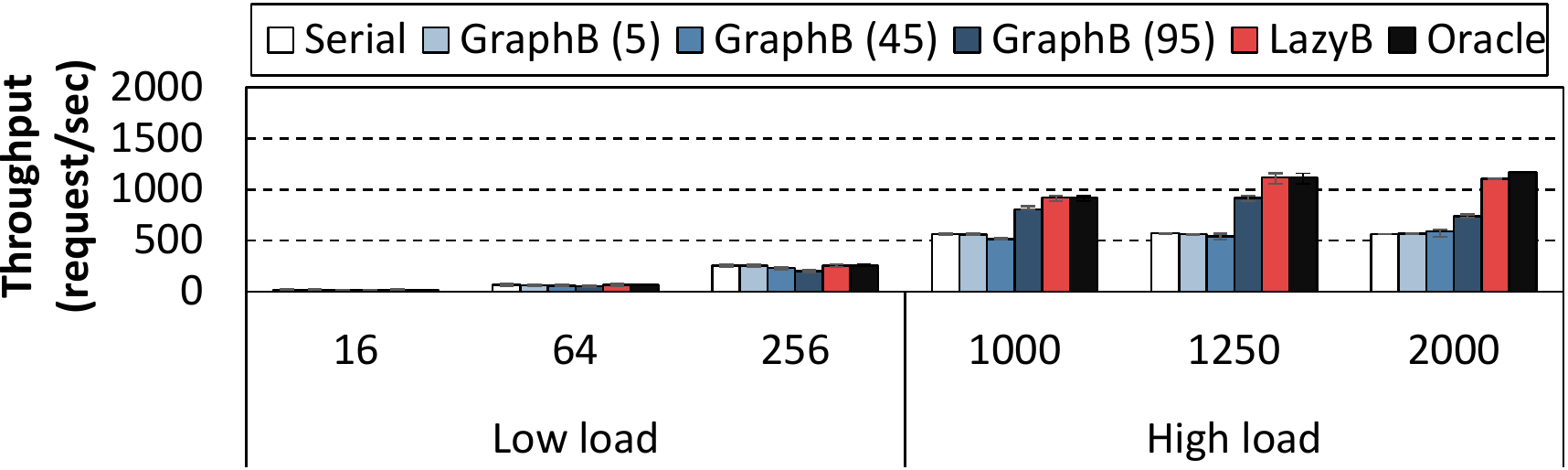}
	\label{fig:throughput_transformer}
}
\vspace{-0.1em} 
\caption{ 
Effect on throughput per query-arrival rate (x-axis, requests/sec). 
}   
\vspace{-0.4em}
\label{fig:eval_throughput}
\end{figure}

\fig{fig:eval_latency} and \fig{fig:eval_throughput} summarize the effect of
different batching policies on average latency and throughput
per inference query-arrival rate (low vs. high traffic).
The error bars represent the
$25$-percentile and $75$-percentile average latency and throughput across
difference simulation runs.  Under low load server traffic conditions, graph
batching consistently performs poorly in terms of both latency and throughput. This is
expected as graph batching needlessly stalls inputs from
execution, despite	having little batching opportunities under low load (especially for large batching
		time-window configurations such as 
		\texttt{GraphB(95)}). 
Consequently, graph batching experiences significantly
longer average latency even compared to \texttt{Serial}, which spends much less time waiting to be issued for execution when
the server is lightly loaded. 
For high loads, graph batching performs better than
\texttt{Serial} as it can amortize the cost of batch collection latency and enhance
throughput. Nonetheless, the statically configured batching time-window fails
to balance latency and throughput and no single graph batching configuration
performs robustly across all applications or server loads. 

\begin{figure}[t!] 
\centering
\includegraphics[width=0.485\textwidth]{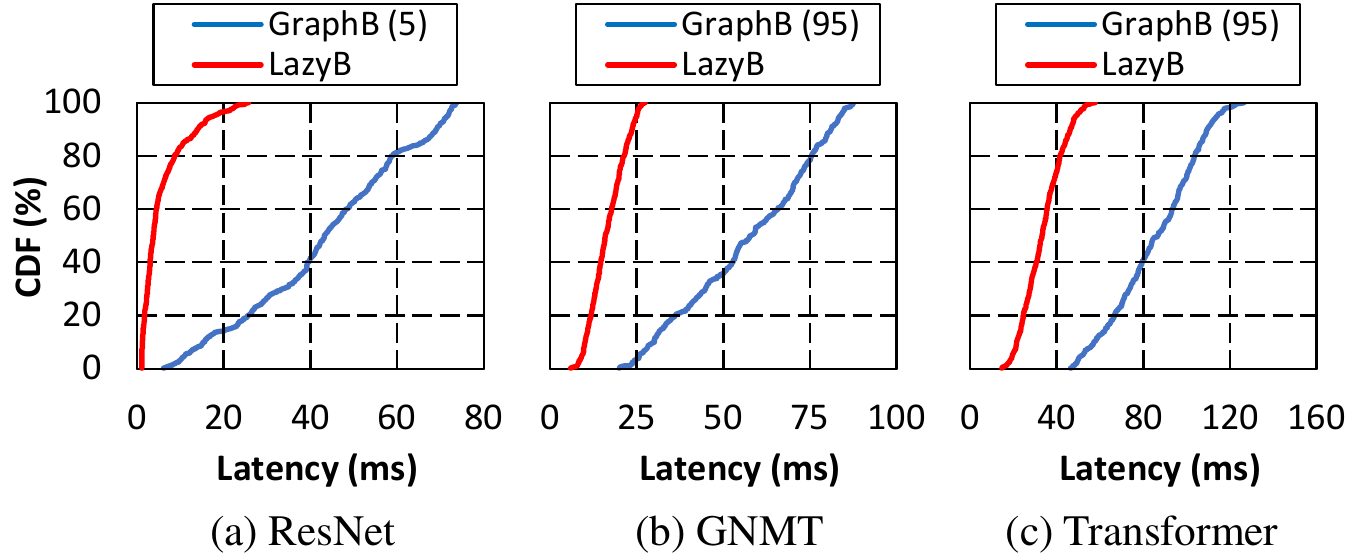}
\vspace{-0.3em} 
\caption{ 
CDF of inference latency under high load (1K req/sec), showing \lazyb's 
	effectiveness in reducing \emph{tail latency}.
For clarity, we only plot the best performing graph batching configuration
for each workload.
}   
\vspace{-1.3em}
\label{fig:eval_cdf}
\end{figure}

Our \lazyb perform superior than both \texttt{Serial} and all graph batching
configurations as it can adaptively adjust to different query-arrival rates,
							 minimizing the latency in forming batched inputs while still
							 reaping out the benefits of batching for improved throughput.
							 Overall, \lazyb provides $5.3\times$, $2.7\times$, and
							 $2.5\times$ lower latency than the best performing graph
							 batching for ResNet, GNMT, and Transformer, respectively.  At
							 the same time, \lazyb provides similar or even better
							 throughput than the throughput-optimized graph batching,
							 achieving an average $1.1\times$/$1.3\times$/$1.2\times$ improvement
							 than the best performing graph batching solution for
							 ResNet/GNMT/Transformer.  These results highlight the robustness
							 of our \lazyb system, which consistently provides low latency while
							 also achieving the throughput benefits of graph batching. We also
							 illustrate \lazyb's merits using \fig{fig:eval_cdf}, which shows
							 the cumulative density function (CDF) of end-to-end inference
							 latency. Notice how the $99$-percentile latency of \lazyb is
							 consistently much smaller than the best performing graph
							 batching (e.g., $54$ vs. $123$ ms of $99$-percentile latency for
									 Transformer), demonstrating the effectiveness of our
							 SLA-aware slack prediction algorithm in reducing \emph{tail latency}.
							 We now further detail \lazyb's effectiveness on minimizing SLA
							 violations thereby guaranteeing QoS.

\subsection{Effectiveness in Meeting SLA Goals}
\label{sect:eval_sla}

\begin{figure}[t!] 
\centering
\subfloat[ResNet]{
\includegraphics[width=0.485\textwidth]{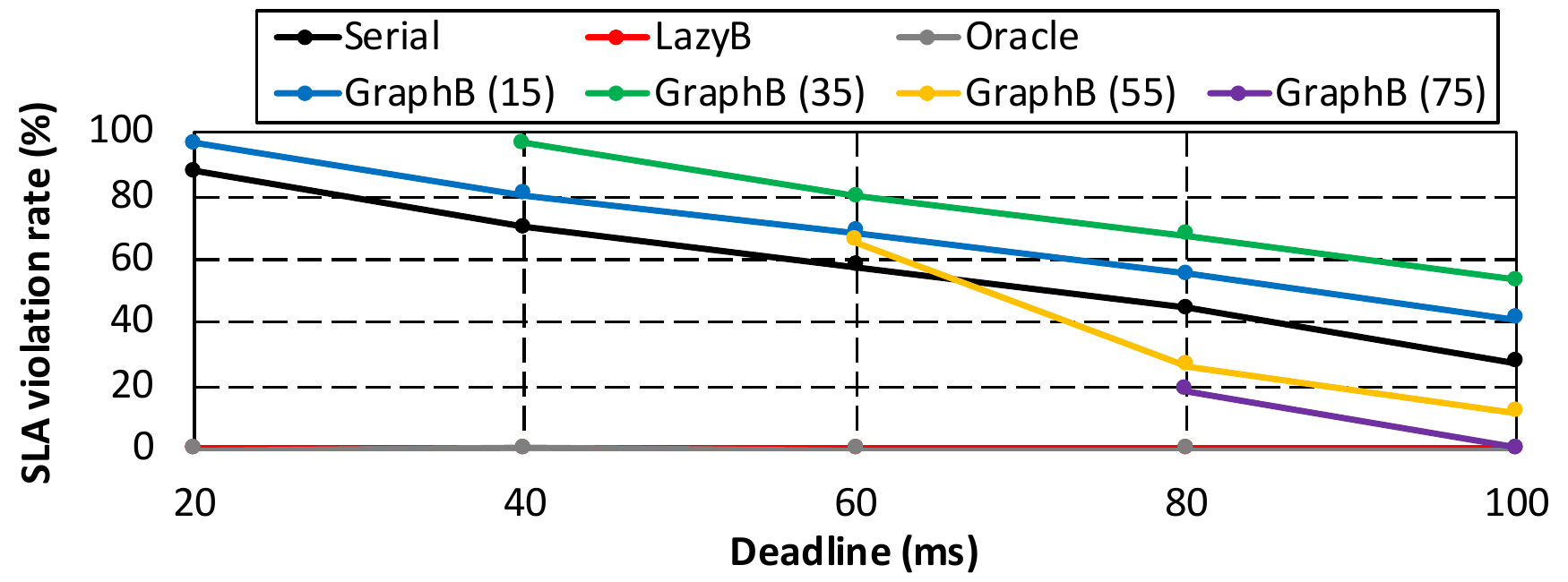}
	\label{fig:sla_resnet}
}
\vspace{-0.1em}
\subfloat[Google Neural Machine Translation (GNMT)]{
	\includegraphics[width=0.485\textwidth]{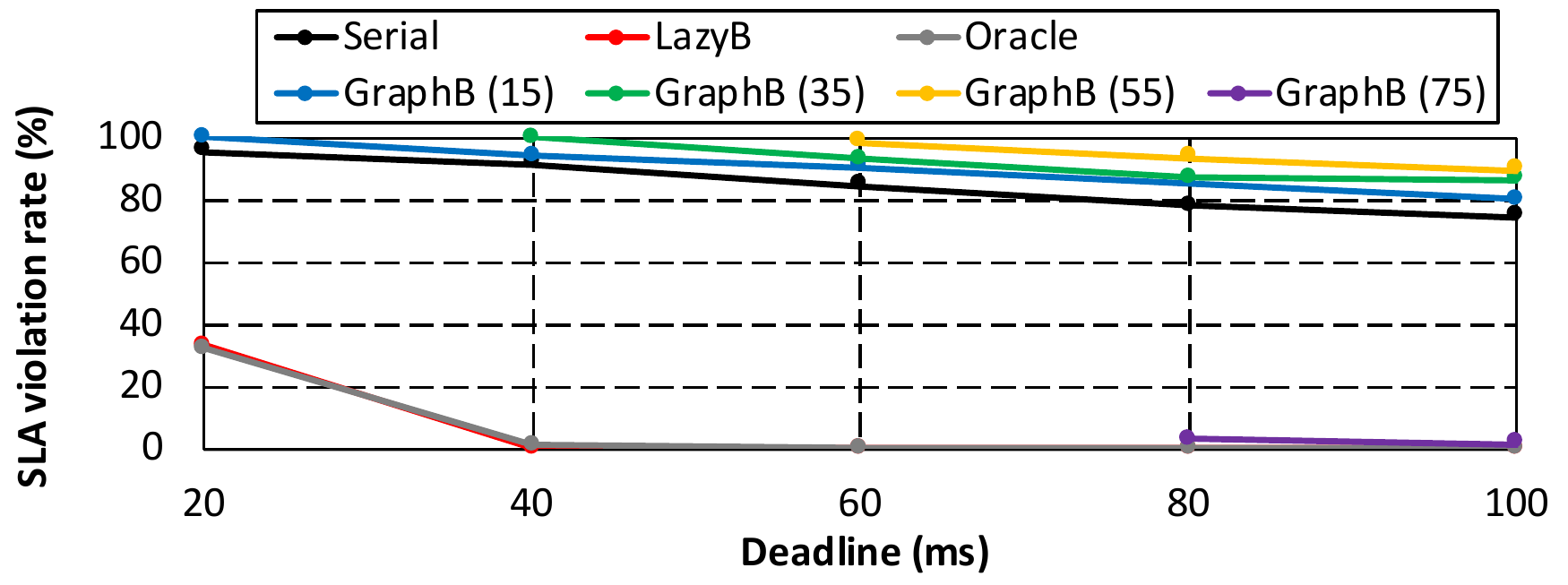}
	\label{fig:sla_gnmt}
}
\vspace{-0.1em}
\subfloat[Transformer]{
	\includegraphics[width=0.485\textwidth]{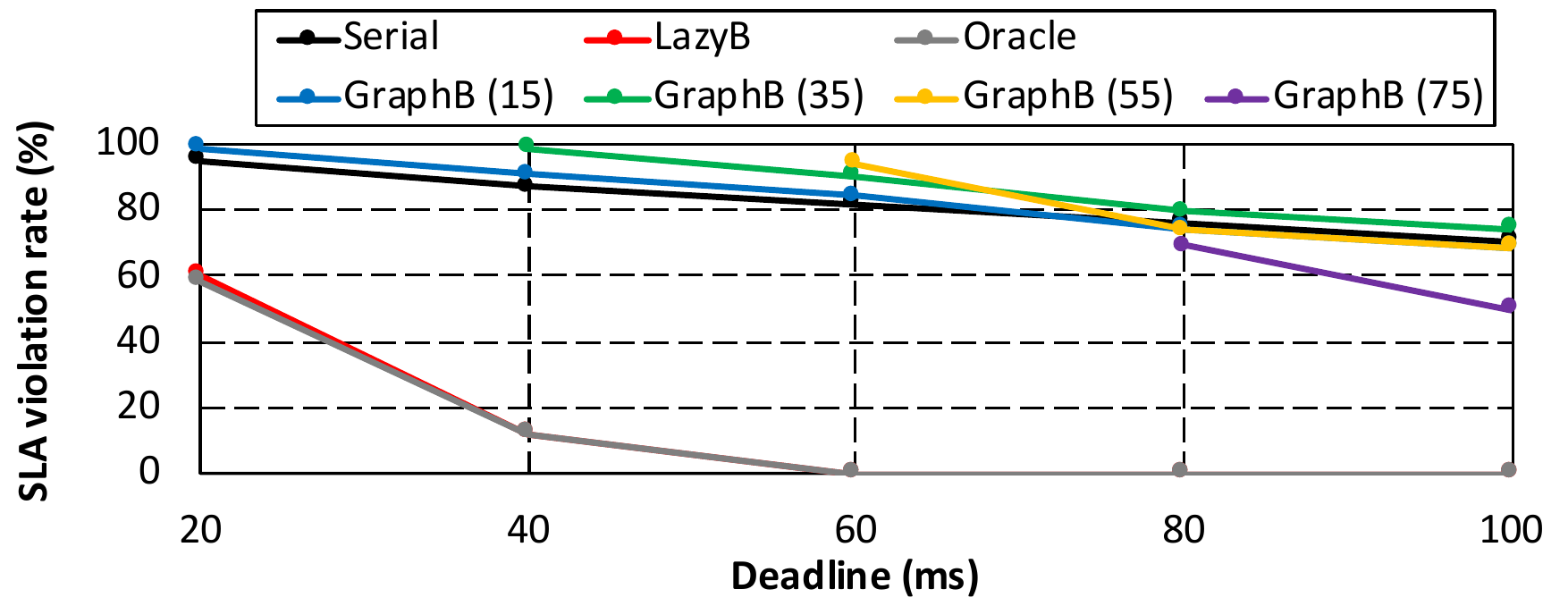}
	\label{fig:sla_transformer}
}
\vspace{-0.2em} 
\caption{ 
SLA violation rate as a function of batching policy and
	SLA deadline (x-axis). The query-arrival rate
	is set to a high load ($1$K req/sec) to stress test a batching policy's
ability to minimize SLA violations (i.e., studying SLA 
under a low query-arrival rate is meaningless because none will violate the SLA).
	We omit plotting impractical data points for brevity (e.g., it does not
			make sense to configure the batching time-window at $75$ ms when SLA deadline is $40$ ms).
    As a SLA deadline increases, from left to right in the x-axis, the violation rate monotonically decreases for all policies.
}   
\vspace{-1.3em}
\label{fig:eval_sla}
\end{figure}

\lazyb's performance is sensitive to the effectiveness of our slack prediction
algorithm, which is dependent on the SLA target value specified per each
model deployment scenario.  Unfortunately, an ML application's SLA deadline target numbers 
are vendor-specific, proprietary information not readily accessible. To
quantify how well our \lazyb scheduler minimizes SLA violations, we sweep the
SLA target value ($SLA_{target}$ in \eqn{eqn:inf_latency}) and measure the
fraction of SLA violated inference requests as a function of different batching
policies. As shown in \fig{fig:eval_sla}, graph batching experiences severe
SLA violations even when the SLA target is set up loosely (e.g., even at SLA
		target of $100$ ms, two-thirds of graph batching configurations experience more than 
		$50\%$ violations).
\lazyb achieves \emph{zero} SLA violations unless the SLA target is set below 
$20$/$40$/$60$ ms for ResNet/GNMT/Transformer, demonstrating its robustness and efficiency even under
such tight SLA constraints. What is also noteworthy is that \lazyb is highly
competitive even when compared against \texttt{Oracle}, which shows the cost-effectiveness
of our lightweight slack prediction algorithm.

\subsection{Sensitivity}
\label{sect:eval_sensitivity}

\begin{figure}[t!] 
\centering
\subfloat[Latency]{
\includegraphics[width=0.485\textwidth]{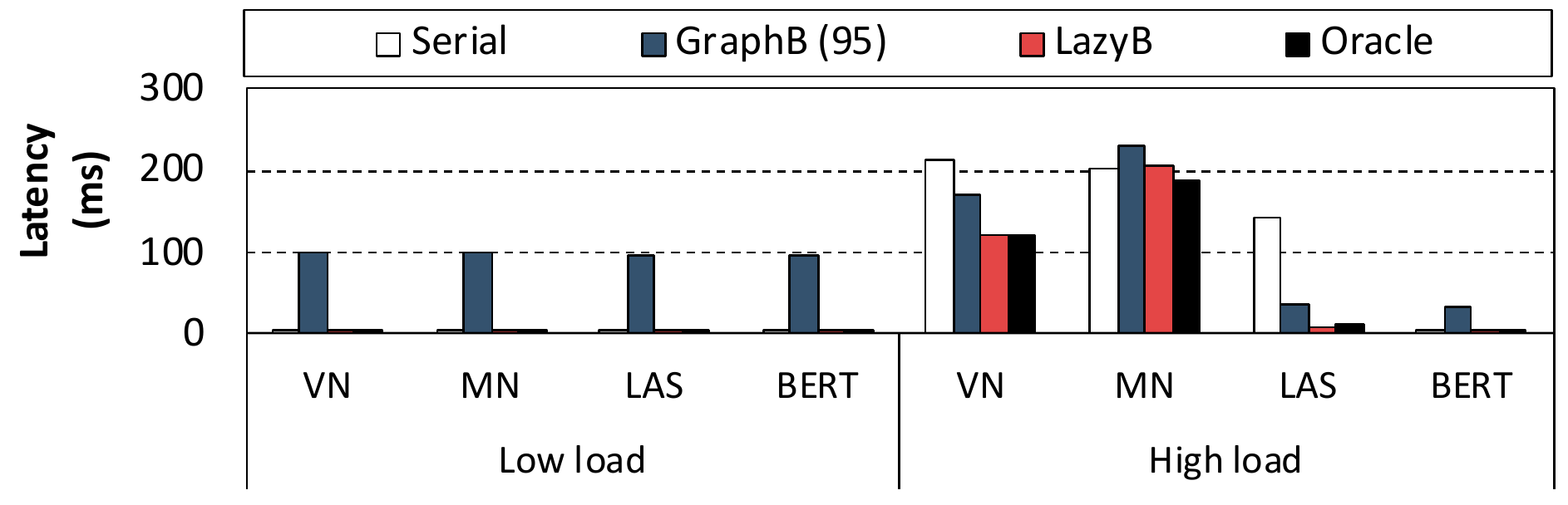}
	\label{fig:app_sensitivity_latency}
}
\vspace{-0.1em}
\subfloat[Throughput]{
	\includegraphics[width=0.485\textwidth]{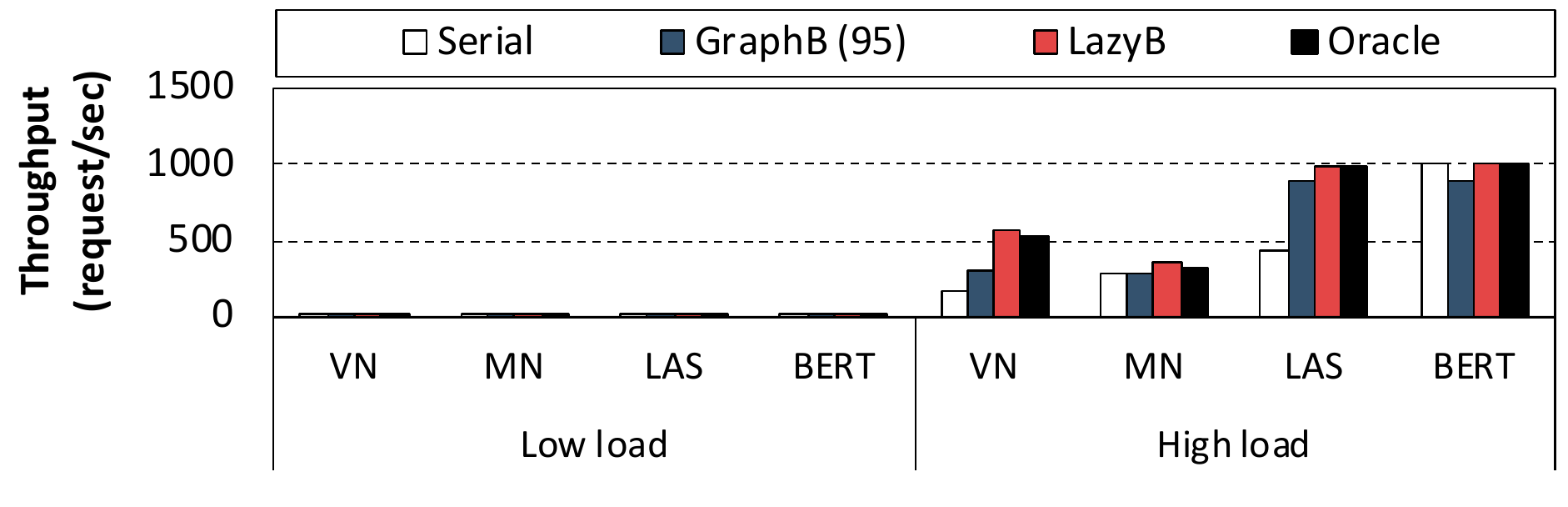}
	\label{fig:app_sensitivity_throughput}
}
\vspace{0em}
\subfloat[SLA violation rate]{
	\includegraphics[width=0.485\textwidth]{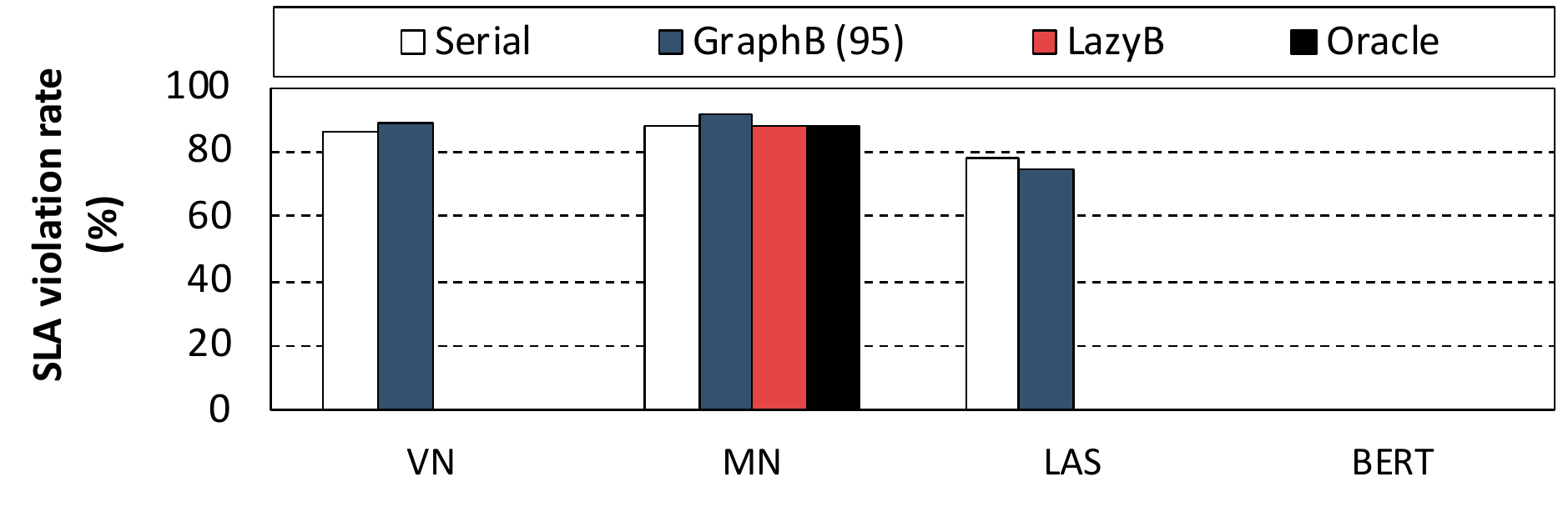}
	\label{fig:app_sensitivity_sla}
}
\vspace{-0.2em} 
\caption{ 
\lazyb sensitivity to other benchmarks. Due to space constraints, we only show
	two datapoints under the low/high
	load in (a,b),  assuming $16$/$1000$ requests/sec, respectively. Similarly, the SLA
	violation rate in (c) summarizes our evaluation under high load ($1000$ requests/sec) where we report
	the average violation rate as a single result when sweeping the SLA deadline from
	$20$ to $100$ ms. BERT's short end-to-end latency renders the assumed $20$-$100$ ms
	SLA deadline to not cause any SLA violations even under \texttt{Serial}. Regardless,
\lazyb significantly improves latency and throughput under this workload.
}   
\vspace{-1.45em}
\label{fig:sensitivity}
\end{figure}

{\bf LazyBatching robustness to other ML applications.} \fig{fig:sensitivity} summarizes the
effect of \lazyb on (a) reducing latency, (b) improving throughput, and
(c) reducing SLA violations, over the four additional benchmarks, VGGNet (VN),
	MobileNet (MN), Listen-Attend-and-Spell (LAS), and BERT. As depicted, our \lazyb
		remains highly robust across a diverse range of applications, achieving
		an average $1.5\times$, $1.3\times$, and $2.9\times$ improvement
		in latency, throughput, and SLA satisfaction, respectively.

{\bf Estimated unrolled sequence length of dynamic DNNs.} \lazyb utilizes the
$dec\_timesteps$ value for estimating dynamic DNN's graph-wide latency
(\algo{algo:graph_latency}).  Under our evaluation setting, choosing a small
$dec\_timesteps$ value leads to an \emph{optimistic} prediction of end-to-end
latency, which increases the estimated slack time and eventually the number of
SLA violations.  For instance, while \lazyb with $dec\_timesteps$=$32$
timesteps (i.e., our default configuration with $N$=$90\%$ coverage) achieves
\emph{zero} SLA violations under an SLA target deadline of $60$ ms, having
$dec\_timesteps$ set to $10$ timesteps ($N$=$16\%$ coverage) leads to an
average $36\%$ SLA violation for Transformer.  Nonetheless, we observe that
\lazyb's performance remains robust as long as $dec\_timesteps$ is sufficiently
large enough to overprovision graph-wide latency thus reducing estimated slack
time.

{\bf Model-allowed maximum batch size.} Prior sections assumed that graph batching's maximum batch size
is set to $64$. When graph batching's maximum batch size is changed to $16$ and $32$, \lazyb achieved
an average $12\times$/$14\times$ latency reduction, and $1.3\times$/$1.3\times$ improved throughput, respectively.

\begin{figure}[t!] 
\centering
\subfloat[Latency]{
	\includegraphics[width=0.485\textwidth]{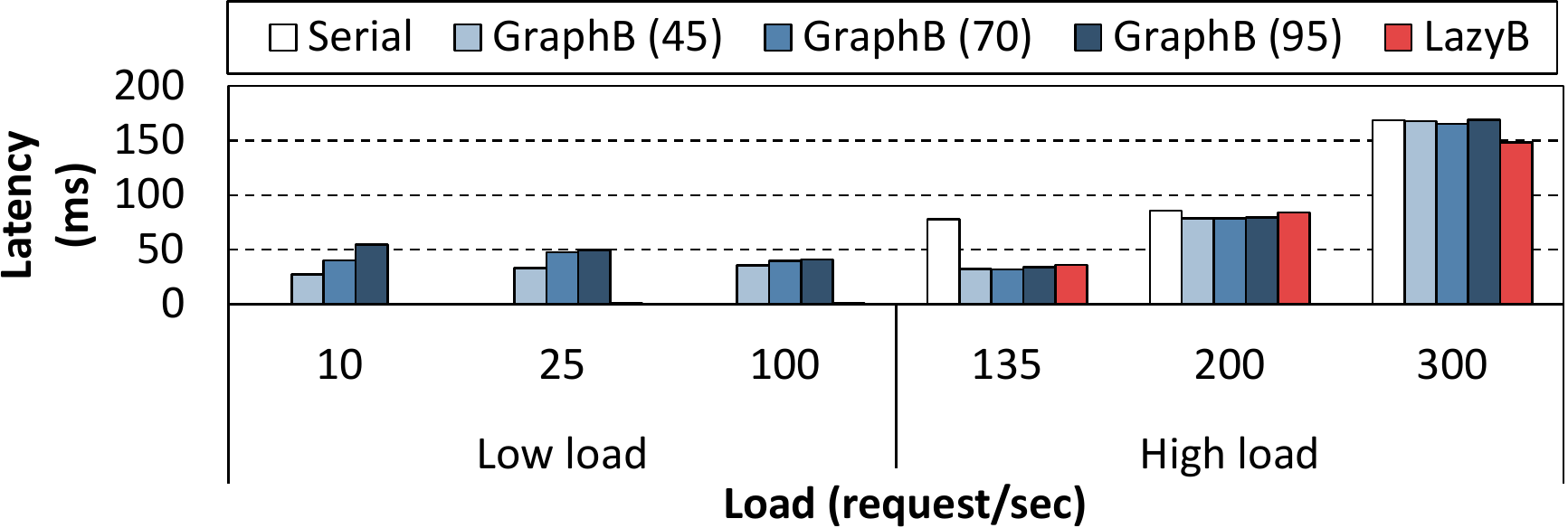}
	\label{fig:gpu_sensitivity_latency}
}
\vspace{-0.1em}
\subfloat[Throughput]{
	\includegraphics[width=0.485\textwidth]{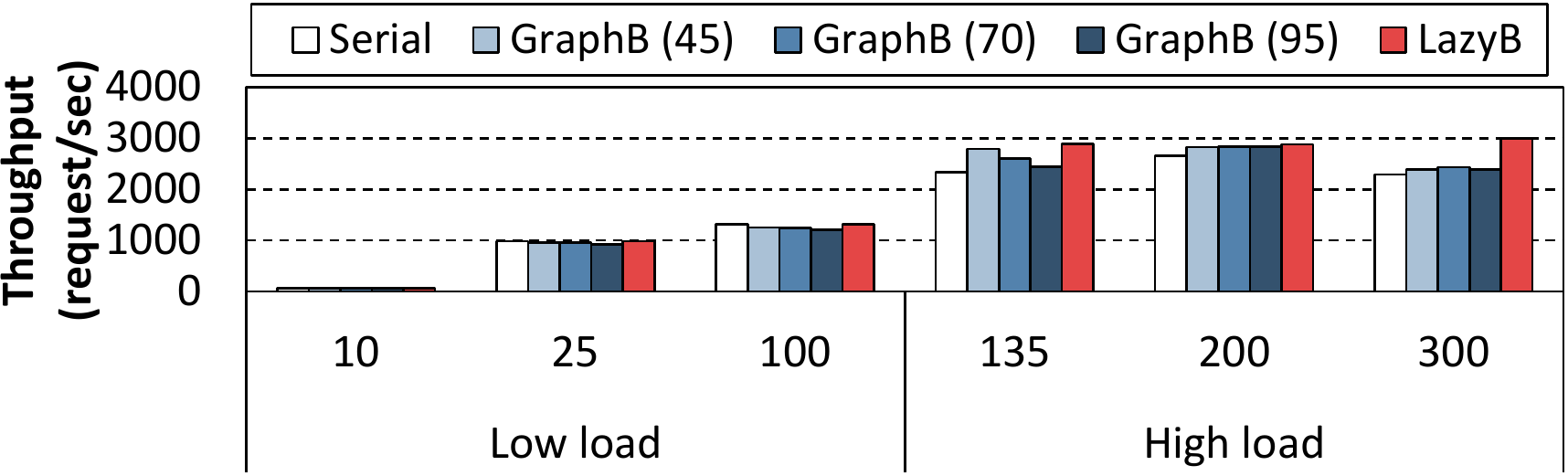}
	\label{fig:gpu_sensitivity_throughput}
}
\vspace{0.5em}
\subfloat[SLA violation rate]{
	\includegraphics[width=0.485\textwidth]{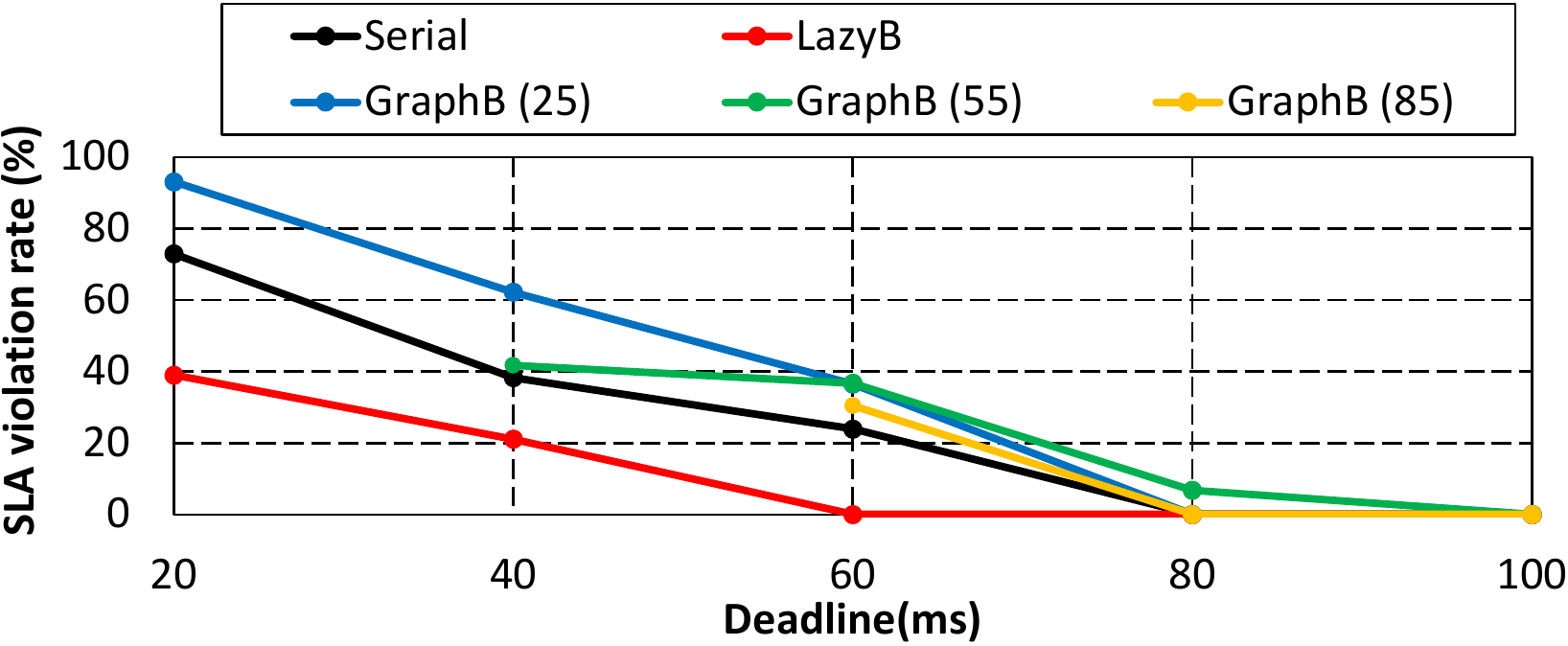}
	\label{fig:gpu_sensitivity_sla}
}
\vspace{-0.2em} 
\caption{ 
LazyBatching sensitivity to GPU-based inference systems. Due to space constraints, we 
only	present detailed results for Transformer, assuming the same evaluation methodology in \sect{sect:eval_latency} and \sect{sect:eval_sla}.}
\vspace{-1.55em}
\label{fig:sensitivity_gpu}
\end{figure}

{\bf Alternative machine translation scenarios.} Our study assumed an English-to-German machine translation pair
as our default evaluation setting, but the effectiveness of \lazyb remains intact for alternative language translation pairs (e.g., Russian-to-English,
		English-to-French, $\ldots$).

{\bf LazyBatching for GPU-based inference systems.} This subsection so far assumed
an NPU-based inference system. We now discuss the applicability and robustness of
\lazyb for GPU-based inference systems. We designed a proof-of-concept software
prototype that is implemented on top of NVIDIA CUDA 10.1 and cuDNN $7.0$. Our
software framework models both the baseline graph batching and our proposed
\lazyb system and the experiments are conducted over an NVIDIA Titan Xp. 
Compared to graph batching, \lazyb provides an
average $1.4$$-$$56\times$ improvement in latency while still achieving
competitive system throughput. In terms of QoS, \lazyb reduces the number of
SLA violations by $1.3\times$ (\fig{fig:sensitivity_gpu}). Overall, \lazyb
shows robustness to GPU-based systems.

{\bf LazyBatching for ``co-located' ML model inference.} Co-locating
	multiple models within a ML inference server helps improve the server's
		overall utilization and therefore its total-cost-of-ownership. To clearly
		separate out the benefits of \lazyb from the advantages coming from
		co-location, we have so far assumed that a single model is deployed within
		the server. To quantify the efficacy of \lazyb under model co-location, we
		follow the methodology employed by Choi et al.~\cite{prema} to
		implement a model inference server supporting model co-location.
		Incorporating \lazyb under co-located ML inference server is
		straightforward. Whenever a new request is received, our scheduler
		examines whether lazily batching this request will violate
		the SLA of the currently on-going requests of co-located ML models,
    which is used to determine batchability.
    We implement our proposal and confirm that \lazyb provides
		an average $2.4\times$/$1.8\times$ improvement in latency and throughput
		than baseline graph batching when when four
		models are co-located.  

		\subsection{Implementation Overhead}
\label{sect:overhead}

As detailed in \sect{sect:lazyb_arch}, \lazyb is based on the node-level DNN
execution model, a property existing ML frameworks and runtime libraries are
already founded upon.  As the stack based batch status tracking process is
purely done in software and the task preemption and context switching is
conducted in node execution boundaries (i.e., layer boundaries), there is no
hardware modifications required to implement \lazyb.  As \lazyb chooses the
node at the top of the stack (i.e., batch state table) for scheduling, the
scheduling computational complexity is \texttt{O(1)} and is thus negligible.  In terms
of memory allocations for batched requests, the required input/output tensors
are allocated upfront to be large enough to accommodate the
model-allowed maximum batch size, which amortizes the runtime
memory management overhead. 
Such
design decision helps remove the memory allocation latency from the critical
path for model inference, a key reason why existing ML inference serving frameworks 
implement
such memory allocation scheme for inference servers.
		 As LazyBatching preempts an on-going batch at the end
of a layer's execution, the output activations are stored into
DRAM, obviating the need for checkpointing intermediate data. As
such, the latency overhead of preemption itself is negligible under
\lazyb.
While we
were not able to implement our software prototype on top of a real NPU hardware
(due to limited availability of NPUs with customizable software frameworks), we
confirm through our software GPU prototype implementation that \lazyb can
readily be implemented on top of existing hardware/software stack.

%% file: tex/discussion.tex
\section{Related Work}
\label{sect:discussion}

While there has been lots of interest in designing energy-efficient NPU architectures 
for training and inference in isolation~\cite{diannao,dadiannao,shidiannao,pudiannao,du:2015:micro,minerva,dnn_pim_reram,eyeriss,cambricon,isacc,neurocube,redeye,tabla,dnnweaver,intel:2017:fpl,gao:2017:tetris,intel:2018:fpga,rhu:2016:vdnn,song:2015:eie,cnvlutin,cambriconx,stripes,bitpragmatic,intel:2017:icassp,intel:2017:fpga,whatmough:2017:isscc,whatmough:2017:hotchips,scnn,bittactical,rhu:2018:cdma,tpu1,kwon:2019:disagg,mcdla,mcdla:cal,tensordimm,neummu,choi:2020:prema,centaur:hwang},
   little attention has been paid in \emph{how} the ML inference server
   collects the batched inputs to feed it into the NPUs.
A few recent literature advocates the need for 
optimizing the batching system in ML. GrandSLAm~\cite{grandslam} explores dynamic
batching for ML application constructed using \emph{microservices}~\cite{amazon_microservice}.
However, unlike \lazyb's fine-grained, layer-wise batching, GrandSLAm conducts batching at the
microservice routine granularity, similar to the coarse-grained, baseline graph-level batching.
PipeDream~\cite{pipedream} exploits \emph{batch-level parallelism} of training to propose an  inter-batch, pipelined execution among 
multiple GPUs. PipeDream's partitioned, inter-batch execution
of different layers bears some similarity to layer-wise execution of \lazyb, but 
	 the scope of this work and the proposed solution 
	 drastically differ against \lazyb. The closest to our work is cellular
	 batching~\cite{cellular_batching}, which we compare and contrast in \sect{sect:limits_cellular_batching} (recall that cellular batching performs identically to baseline
under our workloads). Overall, the key
	 contributions and insights delivered with \lazyb is orthogonal to the aforementioned prior studies.

%% file: tex/conclusion.tex
\section{Conclusion}
\label{sect:conclusion}

While enabling high throughput is a primary design objective in ML
training systems, making sure that the end-user experiences low latency
with high QoS is a fundamental requirement for cloud ML inference.
This paper introduces \lazyb, an intelligent batching system that dynamically 
adjusts the level of batching to meet latency, throughput, and SLA requirements.
Compared to the baseline graph batching, \lazyb provides an average $15\times$, $1.5\times$, 
				 $5.5\times$ improvements in terms of user-responsiveness, throughput, and SLA satisfaction, respectively.
